\documentstyle[11pt,aaspp4]{article}

\def\lae{\mathrel{<\kern-1.0em\lower0.9ex\hbox{$\sim$}}}
\def\gae{\mathrel{>\kern-1.0em\lower0.9ex\hbox{$\sim$}}}

\slugcomment{Accepted to the Astrophysical Journal, July 10 issue}
 
\lefthead{C\^ot\'e, Marzke \& West}
\righthead{Globular Clusters in Elliptical Galaxies}
 
\begin{document}
 
\title{The Formation of Giant Elliptical Galaxies and Their Globular Cluster Systems}
 
\author{{\bf Patrick C\^ot\'e}\altaffilmark{1}}
\affil{Dominion Astrophysical Observatory, Herzberg Institute of Astrophysics,
    National Research Council of Canada, 5071 W. Saanich Road, Victoria, BC, V8X 4M6, Canada,
and \\
California Institute of Technology,
Mail Stop 105-24, Pasadena, CA 91125 \\ pc@astro.caltech.edu}

\author{{\bf Ronald O. Marzke}\altaffilmark{2}}
\affil{Dominion Astrophysical Observatory, Herzberg Institute of Astrophysics,
    National Research Council of Canada, 5071 W. Saanich Road, Victoria, BC, V8X 4M6, Canada,
and \\
Observatories of the Carnegie Institution of Washington, \\
813 Santa Barbara Street, Pasadena, CA 91101 \\ marzke@ociw.edu}

\altaffiltext{1}{Sherman M. Fairchild Fellow}
\altaffiltext{2}{Hubble Fellow}
 
\and

\author{\bf Michael J. West\altaffilmark{3}}
\affil{Department of Astronomy and Physics, Saint Mary's University, \\ Halifax, NS, B3H 3C3, Canada \\
      {\rm west@ap.stmarys.ca}}

\altaffiltext{3}{Also CITA Senior Visiting Fellow, Canadian Institute for 
      Theoretical Astrophysics, University of Toronto, 60 Saint 
      George Street, Toronto, ON M5S 1A7, Canada.}

 
 
\begin{abstract}
We examine the formation of giant elliptical galaxies using their globular cluster
(GC) systems as probes of their evolutionary history. The bimodal
distributions of GC metallicities in such galaxies are often cited as evidence for 
the formation of giant elliptical galaxies through mergers involving gas-rich spirals, 
with the metal-rich GCs forming during the merger process.
We explore an alternative possibility: that these metal-rich clusters
represent the galaxy's intrinsic GC population and that 
the metal-poor
component of the observed GC metallicity distribution arises from the capture 
of GCs from other galaxies, either through mergers or through tidal stripping.
Starting with plausible assumptions for the luminosity
function of galaxies in the host cluster and for the dependence of 
GC metallicity on parent galaxy luminosity, we show using
Monte-Carlo simulations that the growth of
a pre-existing seed galaxy through mergers and tidal stripping is 
accompanied naturally by the capture of metal-poor GCs whose chemical 
abundances are similar to those which are observed to surround
giant ellipticals. 

We also investigate the spatial distribution of GCs in isolated galaxies of 
low and intermediate luminosity and conclude that, at the epoch of 
formation, the GCSs of such galaxies are likely to have been more spatially 
extended than their constituent stars. Thus, the capture of GCs through tidal 
stripping, unlike mergers, does not necessarily conserve GC specific frequency.
Comparisons of model GC metallicity distributions and specific frequencies to those 
observed for the well-studied galaxies NGC 4472 (=M49) and NGC 4486 (=M87), the two
brightest cluster members of the nearby Virgo cluster, show that {\it it is 
possible to explain their bimodal GC metallicity distributions and 
discordant specific frequencies without resorting to the formation of new GCs
in mergers or by invoking multiple bursts of GC formation.}
Finally, we discuss the possibility of using the ratio of the number of metal-poor and 
metal-rich GCs in giant elliptical galaxies as a diagnostic of their merger 
histories. We use this method to derive upper limits on the number of galaxies 
and total luminosity accreted to date by NGC 4472.
\end{abstract}
 
 
 
 
%
 

\section{Introduction}

\bigskip

An understanding of galaxy formation is one of the principal goals of 
modern astrophysics. An important first step in this campaign came with the
demonstration by Toomre (1977) that mergers of late-type galaxies 
often produce end-products which bear a striking resemblance to giant ellipticals (gEs).
In subsequent years, support has grown for the notion that mergers play an
important, if not dominant, role in the formation of gE galaxies. Not only
have investigations into the structure and kinematics of low-redshift gEs 
revealed fossil evidence of past mergers (Bender \& Surma 1992; Tremblay \& 
Merritt 1996; Faber et al. 1997), but theoretical models of galaxy formation invariably 
predict the growth of such galaxies through mergers (Kauffmann, White \& Guiderdoni 1993;
Cole et al. 1994; Navarro, Frenk \& White 1995). Particularly 
close attention has been paid to cD galaxies and centrally dominant ellipticals in rich clusters 
since it is widely believed that the formation of
these rare objects is fundamentally connected to the formation of the clusters themselves
($e.g.$, Ostriker \& Tremaine 1975; Hausman \& Ostriker 1978; Merritt 1984, 1985;
Mackie 1992; West 1994).

The globular cluster systems (GCSs) of gE galaxies have 
traditionally provided some of the most stringent constraints on models 
for the formation and evolution of these objects ($e.g.$, Harris 1986). Since GCs 
are the oldest and brightest objects which can be identified in gE galaxies,
they are powerful tools for investigating the
structure and chemical evolution of their host galaxies at the epoch of
formation. Indeed, any model which seeks to explain the formation of gEs
must be able to account for the {\it number}, {\it chemical abundance} and 
{\it spatial distribution} of their associated systems of GCs.

One of the strongest arguments against the view that gE galaxies 
represent the remains of merged spirals ($e.g.$, Toomre 1977) is 
that of van den Bergh (1982). Simply stated, ellipticals have GC specific frequencies
($i.e.$, the number of GCs per unit parent galaxy luminosity; see \S2.3) which are 
two to three times higher than those of spirals and irregulars (Harris 1991). 
Of course, these morphologically disparate galaxies exhibit
very different clustering properties, so the extent to which their
GCSs have been influenced by environment is unclear (see West 1993).
For instance, Forte, Martinez \& Muzzio (1982) suggested
that the high specific frequency of NGC 4486, the centrally dominant
galaxy in the Virgo cluster, might be explained by the capture of GCs
through tidal-stripping of less massive galaxies (c.f. Muzzio 1987).

An additional constraint on galaxy formation models comes from the observed 
distribution of GC metallicities.
Following initial suggestions of non-Gaussian
metallicity distributions based on rather small samples of GCs (Ostrov, Geisler \& Forte
1993; Lee \& Geisler 1993), a flurry of activity has now demonstrated beyond question
that many, and perhaps {\it most}, luminous ellipticals show distinctly bimodal GC
metallicity distributions ($e.g.$, Whitmore et al. 1995; Elson \& Santiago 1996).
The number of gEs having reliable GC metallicity distributions has now
risen to roughly a dozen (see the recent compilation of Forbes, Brodie \& Grillmair
1997; hereafter FBG97). As these authors point out, the mean metallicity of the {\it metal-rich} component
increases with the luminosity of the parent galaxy in a manner which is entirely
consistent with that originally proposed by van den Bergh (1975).
However, the position of the {\it metal-poor} peak shows no strong
correlation with the properties of the host galaxy:
it falls in the range $-1.5 \lae {\rm [Fe/H]} \lae -0.7$ with little or no dependence 
on parent galaxy luminosity.
The challenge for models of gE formation is therefore to explain not just
the origin and size of these chemically-distinct GC populations, but also to understand
why the location of the metal-poor component peak depends so weakly on the luminosity
of the parent galaxy.

One possible explanation for the high specific frequencies and bimodal 
GC metallicity distributions of gEs is that the same mergers which give rise to
the {\it galaxies themselves} also serve as
catalysts for the formation of new GCs (Schweizer 1986; Ashman \& Zepf 1992). In other words, the
populations of metal-rich GCs associated with galaxies such as NGC 4486
(Whitmore et al. 1995; Elson \& Santiago 1996) and NGC 4472 (Geisler, Lee \& Kim 1996) are
interpreted as being the relics of merger-induced GC
formation, whereas the metal-poor components are assumed to constitute the
{\it pre-existing} GCSs. However, as Geisler, Lee \& Kim (1996) point out,
both of these galaxies show bimodal GC metallicity distributions yet have
specific frequencies which differ by nearly a factor of three (Harris 1986), so
it seems unlikely that any {\it single} mechanism, such as merger-induced GC formation,
can explain entirely the properties of their GCSs.
In light of these difficulties, FBG97 have
argued that the chemically-distinct GC populations in gEs are indigenous to their
parent galaxies and are formed in ``two distinct phases of star formation from gas of differing
metallicity". Nevertheless, it is difficult to identify the mechanism responsible
for the separate GC-formation episodes and for delaying the 
onset of the second burst.

In this paper, we show that bimodal GC metallicity distributions in gEs are a natural
consequence of the capture of GCs through mergers and tidal-stripping of other galaxies. 
Our model differs from that of Ashman \& Zepf (1992) in that we identify the {\it metal-rich}
component of the GC metallicity distributions as the gE's intrinsic
GCS; mergers and tidal stripping will lead almost inevitably to a second peak
in the range $-1.5 \lae {\rm [Fe/H]} \lae -0.7$, with the exact position depending
weakly on the shape of the luminosity function in the host cluster. We also
conclude that, contrary to some earlier claims, the capture of GCs through 
tidal-stripping of other cluster galaxies does not necessarily conserve 
specific frequency, particularly if the bulk of the tidally-stripped GCs 
originated in low- and intermediate-luminosity galaxies. 
Finally, we use this model to explain the bimodal GC metallicity distributions 
and discordant specific frequencies of NGC 4472 and NGC 4486 --- the two brightest 
members of the Virgo cluster.

\section{GC Metallicity Distributions in gE Galaxies: Evolution Due to Mergers and Tidal Stripping}

We seek to model the GC metallicity distributions of gE galaxies, 
particularly those located in rich galaxy clusters. Simply stated, 
our goal is to predict the metallicity distribution of GCs acquired 
through mergers or through tidal-stripping of other cluster galaxies. 
To do this we must first consider three factors which together determine the
total number of GCs of a given metallicity
that are present in any ensemble of galaxies:
(1) the number of GCs as a function of parent galaxy luminosity;
(2) the relationship between GC metallicity and parent galaxy luminosity; and
(3) the distribution of galaxy luminosities.
Combining these three relations allows us to predict the metallicity 
distribution of GCs which may be captured by gEs from other galaxies.

\subsection{The ``Zero-Age" ${\overline{{\rm [Fe/H]}}}$-$M_V^i$ Relation}

There have been numerous compilations of mean GC metallicity, ${\overline{{\rm [Fe/H]}}}$, 
and parent galaxy absolute magnitude, $M_V$, in the recent literature. Although it is 
now established that the two parameters are related in a manner which is
consistent with that originally proposed by van den Bergh (1975),
the exact form of the correlation
remains an open question. Harris (1991) and Durrell et al. (1997) have found a 
slope of $d{\overline{{\rm [Fe/H]}}}/dM_V = -0.17\pm0.04$,
whereas Ashman \& Bird (1993) have argued that genuine {\it halo} GCs have similar
${\overline{{\rm [Fe/H]}}}$, irrespective of parent galaxy luminosity. 
Part of this discrepancy arises from the sample definition, as some investigators have chosen 
to exclude the GCSs of massive galaxies such as gEs and luminous spirals from their analyses
since, at the high luminosities which are characteristic of these objects, the 
observed ${\overline{{\rm [Fe/H]}}}$-$M_V$ relation shows large scatter.
 
Here, we attempt to go a step further and reconstruct the {\it initial}, "zero-age" 
${\overline{{\rm [Fe/H]}}}$-$M^i_V$ relation over the entire range of spheroidal galaxy luminosities.
As a result of the detailed observations of GC colors in gEs, we now know that
the scatter at the bright end of the present-day ${\overline{{\rm [Fe/H]}}}$-$M_V$ relation 
reflects the observed bimodality in the gE metallicity distributions.
Using these detailed color distributions, FBG97 showed that
it is the position of the {\it metal-rich} peak which correlates best
with the gE luminosity.
The existence of such a correlation suggests that the metal-rich GCs in these galaxies formed via the
same mechanism as the GCs in galaxies of low- and intermediate-luminosity, and we are thus
led to identify the metal-rich population as the gE's primordial GCS.  
Our identification of the metal-rich GCs as the intrinsic component of the GCSs
in these galaxies may seem odd at first glance, particularly since GCs are believed to 
be among the first objects to have formed in the early universe.
However, analyses of absorption line indices in gE galaxies using stellar population
models suggest that the gas from which these systems formed was likely to be pre-enriched,
and that the galaxies themselves underwent rapid chemical enrichment by massive
stars ($e.g.$, de Freitas Pacheco 1996; Greggio 1997). Likewise, the extremely high
star formation rates found by Pettini et al. (1997) for primeval galaxies
at redshifts of $z \sim 3$ provide additional evidence for rapid
chemical enrichment in such massive galaxies.
Regardless of the exact mechanism responsible
for the early chemical enrichment, Cohen, Blakeslee \& Ryzhov (1998)
have recently provided compelling evidence that it must have
proceeded on a very short time-scale: Balmer-line indices for 150 GCs surrounding
NGC 4486 indicate that the metal-rich GCs have ages of $\sim 13$ Gyr, indistinguishable from
those of their metal-poor counterparts.
 
If the metal-rich GCs are the intrinsic population ($i.e.$, those originally associated with the seed
galaxy which we now observe as a gE), then the zero-age ${\overline{{\rm [Fe/H]}}}$-$M_V^i$
relation must be defined by the correlation between the seed galaxy luminosity and
the position of the metal-rich peak of the GC metallicity distribution.  The latter is an observed
quantity; unfortunately, the seed galaxy luminosity is not.  However, under the assumption
that the total number of GCs in the galaxy population is conserved during the merger process, 
we can establish the range of possible seed galaxy luminosities from the observed distribution
of GC metallicities.  We explore two limiting cases:
first, we posit that {\it all} of the metal-poor GCs were acquired through mergers. In this case,
the observed fraction of GCs in the metal-rich peak is a direct measure of the luminosity of the
initial seed (provided that the captured galaxies have roughly the same 
GC specific frequency; see \S 2.3):
$$L_{\rm init} = L_{\rm final}~{N_{\rm mr} \over {N_{\rm mr} + N_{\rm mp}}}, \eqno{(1)}$$
or equivalently,
$$M_V^i = M^f_V - 2.5\log{\Big ( {N_{\rm mr} \over {N_{\rm mr} + N_{\rm mp}}} \Big )}, \eqno{(2)}$$
where $M_V^f$ is
the final absolute magnitude of the gE galaxy, and $N_{\rm mr}$ and $N_{\rm mp}$
refer to the number of metal-rich and metal-poor GCs, respectively.  Since all three
of these present-day quantities are available from FBG97, we can compute an initial, 
pre-merger ${\overline{{\rm [Fe/H]}}}$-$M_V^i$ relation directly from these observations
(assuming that the dynamical friction time-scale in the merged system is long, and that 
the metal-rich and metal-poor GCs are equally susceptible to destruction; see \S4.3).
Strictly speaking, our ``zero-age" relation is more accurately described as the 
initial ${\overline{{\rm [Fe/H]}}}$-$M_V^i$ relation {\it as it would appear today}
provided the subsequent evolution has been totally passive ($i.e.$, we apply no magnitude 
correction to account for the fading stellar populations in the original galaxies, 
since the same correction would then need to be subtracted when comparing to nearby gEs).

In the opposite limit, we make the extreme assumption that the initial luminosity is the same as
the final luminosity: $i.e.$, that the metal-poor GCs are stripped from neighboring galaxies
without {\it any} accompanying field stars.  This is obviously a firm
upper limit to the seed luminosity and a conservative bound at that; stripping will inevitably
transfer at least some field stars to the seed galaxy.\altaffilmark{4}\altaffiltext{4}
{One might also consider a third case: that field stars are actually accreted {\it more}
efficiently than GCs.  However, the fact that no GCSs are observed to be less extended
that the underlying stellar halo makes this possibility seem very remote.}
 
The resulting range of possible ${\overline{{\rm [Fe/H]}}}$-$M_V^i$ relations 
is shown in Figure 1.  For the gEs, we plot
the eleven gEs listed in FBG97. For each galaxy, we adopt the mean metallicity for the
metal-rich GC population reported in FBG97 and assume an uncertainty in ${\overline{{\rm [Fe/H]}}}$ 
of $\pm0.2$ dex.
The modest corrections derived from Equation 2 ($e.g.$, $M_V^i - M_V^f$ $\simeq$ 1 mag)
are shown as the horizontal arrows in Figure 1, where
we have assumed an uncertainty of $\pm0.3$ mag in $M_V^i$ for each galaxy.
 
Because dwarf galaxies are less likely to have merged with smaller systems, they provide a more direct
view of the original ${\overline{{\rm [Fe/H]}}}$-$M_V^i$ relation.
A search of the literature revealed nine dwarf elliptical galaxies 
for which there exist data on the chemical abundance of their GCSs.  
For these galaxies, we have plotted the raw values of ${\overline{{\rm [Fe/H]}}}$ and 
$M_V$ in Figure 1.
For each galaxy, we have taken the best available estimate for 
the metallicity of each individual GC. For a few of the galaxies, there exist multiple 
determinations of the metallicities of some GCs; in such cases, we give 
preference to spectroscopic measurements.
We then calculate the mean GC metallicity ${\overline{{\rm [Fe/H]}}}$,
the intrinsic metallicity dispersion ${\sigma}_i$, and their respective uncertainties
for each galaxy. 
 
We also include in our sample NGC 1380: the only lenticular (S0) galaxy in
which accurate metallicities have been measured for a large number of GCs 
(Kissler-Patig et al. 1997). Like most gE galaxies,
NGC 1380 shows a bimodal GC metallicity distribution.
Following Kissler-Patig et al. (1997), we adopt ${\overline{{\rm [Fe/H]}}} = 0.15\pm0.5$,
$N_{\rm mr} = 382\pm20$ and $N_{\rm mp} = 182\pm20$.
 
Our sample of 21 galaxies is listed in Table 1,
whose columns record the name, morphological type and
absolute visual magnitude for the parent galaxy of each GCS, the total number of GCs
having measured metallicities, and the corresponding mean GC metallicity and intrinsic
dispersion; references are given in the final two columns. 
Since it is clear that the relationship between ${\overline{{\rm [Fe/H]}}}$ 
and $M_V^i$ is non-linear, we parameterize the ${\overline{{\rm [Fe/H]}}}$-$M_V^i$ relation
in the form
$${\overline{{\rm [Fe/H]}}} = a_0 + a_1M_V^i + a_2{M_V^i}^2. \eqno{(3)}$$
We then find the constants $a_0$, $a_1$ and $a_2$ which minimize the goodness-of-fit statistic
$$\chi^2(a_0, a_1, a_2) = {\sum_{j=1}^{N}} { (~\overline{[{\rm Fe/H}]}_j - a_0 - 
a_1M^i_{V,j} -a_2{M^{i~2}_{V,j}})^2 \over
{\epsilon}_{\rm {\overline{[{\rm Fe/H]}}}}^2 + (a_1 + 2a_2M^i_{V,j}){\epsilon}_{\rm {\overline{[{\rm Fe/H]}}}}^2} \eqno
{(4)}$$
where ${\epsilon}_{\rm {\overline{[{\rm Fe/H]}}}}$ is the uncertainty in the mean
GC metallicity, and we have made use of the fact that 
$${d{\overline{{\rm [Fe/H]}}} \over dM_V^i} = a_1 + 2a_2M_V^i. \eqno{(5)}$$
Minimizing Equation 4 yields 
$a_0 = 2.31$, $a_1 = 0.638$ and $a_2 = 0.0247$. The resulting ${\overline{{\rm [Fe/H]}}}$-$M_V^i$
is shown as the dotted line in Figure 1; the dashed line indicates the relation found
without the correction given by Equation 2.  The difference between the two fits
is never large; we show in \S 3 that our conclusions depend very little on these
small variations in the ${\overline{{\rm [Fe/H]}}}$-$M_V^i$ relation.
 
Finally, we investigate the possibility that the intrinsic dispersion in 
GC metallicity also depends on the luminosity of the parent galaxy.
Based on the nine galaxies in Table 1 which have reliable estimates for ${\sigma}_{i}$, we find a
weighted mean dispersion of $\overline{{\sigma}}_{i} = 0.28\pm0.05$ dex with no significant
evidence for a correlation between ${\sigma}_{i}$ and $M_V$. The increased metallicity 
dispersion among luminous galaxies noted by previous investigators ($e.g.$, Harris 1991)
is likely to be a consequence of their bimodal GC metallicity distributions which went 
unnoticed for many years due to the small sample sizes and the limited metallicity 
sensitivity of the color indices used in early photometric surveys (see the 
discussion in Geisler, Lee \& Kim 1996).

\subsection{The Initial Luminosity Function of Cluster Galaxies}

Galaxies in clusters span an enormous range in luminosity and mass. The present-day 
galaxy luminosity function (LF) is determined both by the cosmological conditions when galaxies
began to form and by subsequent evolution of the galaxy population. 
In the case of an $\Omega=1$, cold-dark-matter universe, the predicted mass spectrum is steep
at all redshifts; at $z=0$, $dN/dM \propto M^{-2}$ at the low-mass end. 
The corresponding LF follows from the cycle of gas cooling, star
formation, and supernova feedback, none of which (with the possible exception of the first) are 
well understood.  Nevertheless, solutions derived using quite disparate physical assumptions
yield a remarkably consistent picture. Depending mostly on the strength of the feedback,
the derived LFs are always steep at the faint end:
$ -1.4 \le d\log{N}/d\log{L} \le -2.0$ ($e.g.$, Kauffmann, Guiderdoni \& White 1994, Cole et al. 1994, Baugh et al. 1996).
 
Unfortunately, our theoretical understanding of galaxy formation remains incomplete, and direct
extrapolation of the redshift-dependent LFs is still our
best hope for reconstructing the initial LF.
Of the many analytic expressions proposed to describe
the observed galaxy LF, $\phi(L_V)$,
the Schechter (1976) function is the most widely used:
$${\phi}(L_V)dL_V = {\phi}^* \Big ( {L_V / L_V^*} \Big )^{\alpha} 
{\rm exp} \Big ( -{L_V / L_V^*} \Big )d({L_V / L_V^*}) \eqno{(6)}$$
where ${\phi}^*$ a parameter related to the number of galaxies per unit volume, $L_V^*$ is a 
characteristic luminosity (measured for our purposes in the $V$ band)
and ${\alpha}$ is a parameter which determines the relative number 
of faint and bright galaxies. 
 
Enormous effort has gone into measuring the galaxy LF both in the field and in clusters
($e.g.$, see the reviews by Binggeli, Sandage \& Tammann 1988 and Ellis 1997). 
Virtually all studies 
agree that at low redshift, 
the LF turns over sharply near $L_V^*$ and is essentially flat (in $\log \phi - \log L_V$)
down to approximately three magnitudes below $L_V^*$.
Fainter than this, however, there is
considerable disagreement over the LF slope.
For instance, Marzke et al. (1994) find evidence for a
steep upturn of $\alpha \simeq -1.9$ at the faint end of the field LF, while
Ellis et al. (1996) claim that the LF remains flat to at least four and a half 
magnitudes fainter than $L^*$.
Recently, Loveday (1997) has suggested that a better parameterization of the field LF is
a double Schechter-function with a very steep faint-end slope of $\simeq -2.8$.

Several recent studies of the cluster LF have produced generally similar results.
For example, in their study of the Coma cluster, Biviano et al. (1996) have shown that a two-component LF, 
consisting of a Gaussian at the bright end and a steep Schechter function at the faint end, 
closely resembles the observed LF.
Smith, Driver \& Phillips (1997), Wilson et al. (1997), Lopez-Cruz et al. (1997) and 
Trentham (1997) find similar behaviour in several rich clusters over the redshift range
$0 \lae z \lae 0.2$.  However, these authors disagree on the question of how cluster LFs
have evolved with time; for instance, Smith, Driver \& Phillips (1997) find little difference between clusters
at $z = 0$ and $z = 0.2$, while Wilson et al. (1997) claim that the galaxy colors suggest
the presence of a fading population of dwarf irregulars whose star formation was truncated soon after
$z \simeq 0.2$.  

Given the uncertainty in both field and cluster LFs at early epochs, we choose to
explore a range of possibilities for the initial LF.
We are particularly interested in the initial LF
in clusters, since it is bright cluster ellipticals that we trying to understand.  Given the observations,
it seems likely that the present-day LF in, for example, the Virgo cluster represents
a lower bound on the {\it initial} faint-end slope: $\alpha=-1.2$.  As an upper bound, we choose a single
Schechter function with the slope of the steep component observed in many low-redshift clusters:
$\alpha=-1.8$. Such a steep LF might represent a brighter analog of the steep (perhaps faded)
dwarf LF observed at low redshift.  
Because we find that single Schechter functions covering this range
in $\alpha$ yield a range of metallicity distributions which encompasses those produced by two-component
LFs, we restrict our analysis to the single Schechter function with $-1.8 \le \alpha \le -1.2$.
 
\subsection{Choice of GC Specific Frequency}

During interactions with other galaxies, the number of GCs available for 
capture by a gE galaxy depends on the luminosity of each passing
galaxy. To first order, the number of GCs, $N_{\rm gc}$, scales directly
with absolute magnitude of the parent galaxy. A convenient means of expressing 
the number of GCs per unit galaxy luminosity is the specific frequency, $S_n$, which 
was first proposed by Harris \& van den Bergh (1981),
$$S_n = N_{\rm gc}10^{0.4(M_V + 15)}. \eqno{(7)}$$
For dwarf ellipticals and gEs in rich clusters, $S_n \simeq 5$ with a dispersion
of ${\sigma}(S_n) \simeq 1$ ($e.g.$, Harris 1991).
In what follows, we assume specific frequency to be independent of $M_V$ 
(see Harris 1991 and references therein) and adopt a specific frequency of
${\overline{S_n}} = 5\pm1$ for the interacting galaxies.
In \S3.1, we discuss the implications of this choice of ${\overline{S_n}}$.

\section{Monte Carlo Simulations of GC Metallicity Distributions}

Having specified the ${\overline{{\rm [Fe/H]}}}$-$M_V^i$ relation, the specific 
frequency ${\overline{S_n}}$, and the initial galaxy LF, it is possible to simulate the 
metallicity distribution of GCs captured by a gE galaxy. As discussed 
in \S2.1, we assume that the luminosity of the initial seed galaxy, $L_{\rm init}$,
has grown by the amount $L_{\rm cap}$ as a result of the capture of smaller galaxies. 
Therefore, we calculate $L_{\rm cap}$ for each gE as follows,
$$L_{\rm cap} = L_{\rm final}~{N_{\rm mp} \over {N_{\rm mr} + N_{\rm mp}}}, \eqno{(8)}$$
and select at random a galaxy which, based on the probability 
distribution given by Equation 6, may have merged with the initial gE galaxy.
Since, by definition, the gE galaxy cannot consume an object larger than itself, we 
adopt a maximum luminosity of $L_{\rm init}$ for Equation 6. Likewise, we
assume that the faint-end of the LF is truncated at $L_V = 10^7L_{V,{\odot}}$. 
This is the luminosity of the Fornax dwarf galaxy, the faintest galaxy known to 
harbor its own GCS (Mateo et al. 1991; Harris 1991).
We repeat this process, recording the combined luminosity of the captured galaxies,
$$L_{\rm cap} = {\sum_{k=1}^{N_{\rm cap}}} {L_k}, \eqno{(9)}$$ 
until the condition $$L_{\rm cap} \ge L_{\rm final} -  L_{\rm init} \eqno{(10)}$$ is satisfied.
At this point, the simulations are terminated and the total number of captured galaxies, 
$N_{\rm cap}$, is recorded. 

For each captured galaxy we calculate the total number of associated GCs using Equation 7,
including a random dispersion of ${\sigma}(S_n) = 1$ for both the 
captured and seed galaxies.  A metallicity is assigned randomly to each captured GC from an assumed
Gaussian distribution with mean metallicity, ${\overline{{\rm [Fe/H]}}}$, determined from
Equation 3, and an intrinsic spread of $\overline{{\sigma}}_{i} = 0.28$ dex, irrespective of
the host galaxy luminosity. We account for the scatter in the 
${\overline{{\rm [Fe/H]}}}$-$M_V^i$ relation by including a random dispersion 
of 0.2 dex when assigning a mean GC metallicity to each galaxy (as discussed in \S2.1).
We repeat this process one last time for the initial gE galaxy
and combine the intrinsic and captured GCs to create a final simulated
GC metallicity distribution.

For simplicity, we assume that all galaxies have equal merger probabilities; in reality,
the likelihood of capture may be a function of galaxy mass. For instance, circular orbits 
through an isothermal dark halo decay on a time-scale inversely proportional to the mass, $M$
(Ostriker \& Tremaine 1975; Tremaine 1976).  For a more general distribution of orbits, however,
the situation is more complicated. Cole et al.  (1994) have argued on the basis of the 
N-body/hydrodynamical simulations of Navarro, Frenk \& White (1995) that the merger 
time-scale, $\tau$, shows a relatively weak mass dependence: $\tau \propto M^{-0.25}$. 
Since the exact form of the dependence is still under debate, we have chosen not to 
include an explicit mass dependence in our models.  We simply note that for a constant
mass-to-light ratio, the merger probability derived by Cole et al. (1994) is proportional
to $L^{0.25}$. If this is indeed the correct dependence, then each of the 
LFs considered in the following analysis would need to be corrected by an 
amount $\Delta\alpha = -0.25$ ($i.e.$, an assumed LF slope of 
$\alpha = -1.5$ would correspond to an actual slope of $\alpha = -1.75$).

As an illustration of this method, we now compare the simulated GC metallicity 
distributions to those observed for a well-studied gE galaxy: NGC 4472, 
the brightest member of the Virgo cluster.

\subsection{The GC Metallicity Distribution of NGC 4472}

The GC metallicity distribution of NGC 4472 
represents the best existing dataset for any gE galaxy in terms of 
calibration, metallicity sensitivity and large sample size
(Geisler, Lee \& Kim 1996). This last factor is usually the most severe limitation 
on other datasets, which typically consist of relatively small numbers of GCs. In such cases,
the low-order moments of the GC metallicity distributions (such as the the mean metallicities
of the metal-rich and metal-poor peaks) are reasonably well constrained, but it is 
impossible to model the detailed shape of the distribution function. Recent summaries of bimodal
GC metallicity distributions in gE galaxies can be found in Geisler, Lee \& 
Kim (1996) and FBG97.

As mentioned in \S2.2, choosing the values of $\alpha$ and $L_V^*$ (or, alternatively, $M^*_V$) 
to be used in the simulations is complicated by the possibility that the {\it original} Virgo 
LF differs significantly from the {\it present-day} LF. Sandage, Binggeli \& Tammann (1985) measure
$\alpha = -1.25$ and $m_B^* = 10.6$ for the present-day, composite LF in Virgo,
although the slope of the LF at the faint end depends rather strongly on the sample definition.
For instance, the best-fit LF for dEs and gEs alone has $\alpha = -1.45$.
We have therefore carried out simulations for a range of assumed LF slopes,
bearing in mind that $\alpha = -1.25$ is likely to represent a lower limit on the initial LF slope.
The evolution of $M_V^*$ is expected to proceed
more slowly, so we have combined the present-day estimate of $m_B^* = 10.6$ with the Virgo distance
modulus of $(m-M)_0 = 31.04$ mag (Ferrarese et al. 1996), a Galactic reddening of
$E(B-V) = 0.10$ and a mean galaxy color of ${\overline{(B-V)}} = 0.9$ at $m_B^*$
($e.g.$, Roberts \& Haynes 1994) to find $M^*_V = -21.7$.

Figures 2 -- 4 compare the results of our simulations with the observed GC metallicity 
distribution for NGC 4472 (Geisler, Lee \& Kim 1996). As initial conditions for the 
simulations, we assume $M_V^i = -21.6$ and $M_V^f = -22.7$ for NGC 4472 (see below).
We show four different simulations for each assumed LF slope, chosen to 
constitute a representative sample of simulated GC metallicity distributions.
As a measure of the similarity between
the observed and simulated distributions, we calculate the 
goodness-of-fit statistic
$${\chi}^2_{\rm sim} = {1 \over N_{\rm bin} - 1} {\sum_{i=1}^{N_{\rm bin}}} {({N_{\rm{obs,i}} - N_{\rm{sim,i}}})^2 \over (N_{\rm{obs,i}}
+ N_{\rm{sim,i}}) } \eqno{(11)}$$
for each simulation. The corresponding values of ${\chi}^2_{\rm sim}$ are recorded
in Figure 2 -- 4. 

If we assume that {\it all} of the metal-poor GCs in NGC 4472 were captured in mergers, 
then it is possible to calculate the total number of galaxies consumed by 
NGC 4472.\altaffilmark{5}\altaffiltext{5}
{In \S4.2 we argue that the capture of GCs through tidal stripping has been more 
important in NGC 4486 than in NGC 4472 since the former is located at the bottom of 
the potential well of the main Virgo cluster, close to where the cluster tidal forces reach their maximum.}
Of course, this is possible simply because the capture of even a single GC in a 
merger naturally entails the capture of the entire host galaxy, something which
is {\it not} true of tidal stripping.
The total number of captured galaxies is recorded in each panel of Figures 2 -- 4.
Clearly, the stochastic nature of the merger process strongly influences the final
GC metallicity distribution, as a galaxy may increase its luminosity in any number of 
ways: by swallowing many small dwarfs or just a few large galaxies.
The effect is, however, less pronounced for steep LFs since there are relatively fewer
luminous galaxies, and hence
the initial gE is unlikely
to capture a comparably-bright galaxy.  In such cases, the final simulated GC metallicity
distribution is almost always bimodal.

According to FBG97, the absolute visual magnitude of NGC 4472 is $M^f_V = -22.7$.
Since the the observations of Geisler, Lee \& Kim (1996) indicate that 35\% of the GCs
in NGC 4472 are metal-rich, our estimate for the absolute
magnitude of the captured component in NGC 4472 is $M_V = -21.6$, 
which corresponds to $\sim$ 65\% of its present-day luminosity. The number of captured 
galaxies depends on both ${\overline{S_n}}$ and the assumed LF: for
LF slopes of $\alpha = -1.2, -1.5$ and $-1.8$, the median number of captured 
galaxies found in 1000 Monte Carlo simulations (see below) are 
44$^{+19}_{-16}$(5/${\overline{S_n}}$), 148$^{+50}_{-48}$(5/${\overline{S_n}}$)
and 508$^{+109}_{-107}$(5/${\overline{S_n}}$), respectively. Although the total number
of captures depends very strongly on the assumed LF, the majority of the captured 
galaxies are in most cases low-luminosity systems, each of which contributes 
relatively few GCs. For our best-fit slopes of $\alpha \sim -1.8$, the bulk of the metal-poor
GCs originate in galaxies having absolute magnitudes in the range $-15 \lae M_V^i \lae -19$.

It is worth bearing in mind that such estimates for the number of captured galaxies 
are best interpreted as {\it upper limits} since tidal-stripping may have 
produced some of the observed metal-poor GCs. However, in the absence of large 
variations in ${\overline{S_n}}$ within a single galaxy cluster --- or between 
clusters which are morphologically and dynamically similar --- this technique 
may provide a useful method of comparing gE merger histories.

\subsubsection{Frequency of Bimodality}

It is a remarkable fact that the GC metallicity distributions of virtually all
well-studied gEs to date show some evidence for bimodality. How
frequently do our simulated GC metallicity distributions exhibit such bimodality?
To answer this question, we have generated 1000 simulated GC metallicity distributions
for NGC 4472, assuming LF slopes of $\alpha = -1.2, -1.5$ and $-1.8$.
As before, the simulations have been binned in exactly the same fashion as the observations
of Geisler, Lee \& Kim (1996). For each simulation, we fit single- and double-Gaussian 
distributions to the resulting GC metallicity distribution, and determine the reduced $\chi^2_{\nu}$ for both 
parameterizations.\altaffilmark{6}\altaffiltext{6}{The number of degrees of freedom, ${\nu}$, 
is ten for the single Gaussian and seven for the double Gaussian.}
An $F$-ratio test is then used to determine if the improvement obtained by including
a second component is warranted (Bevington 1969).
In the three left panels of Figure 5, we show histograms of the 
$F$-ratio statistic for 1000 simulated distributions having LF slopes of $\alpha = -1.2, -1.5$ 
and $-1.8$. As expected, bimodality becomes increasingly obvious as LF slope is
increased; for $\alpha = -1.8$, 94\% of the simulated distributions show bimodality.
However, even for rather flat LFs, the majority of the simulations are
bimodal. For instance, for an LF slope of $\alpha = -1.2$, 80\% of the simulated
distributions show statistically significant evidence for bimodality.
It is also worth pointing out that although the simulated GC metallicity
distributions are, in the vast majority of cases, well described by double Gaussians,
for rather steep LFs there are occasions when more than two peaks are evident
in the simulations.\altaffilmark{7}\altaffiltext{7}{Note that Lee \& Geisler (1993) have
suggested that the GC metallicity distribution in M87 shows evidence for {\it three} peaks
based on ground-based photometry for $\simeq$ 400 GC candidates, although
Whitmore et al. (1995) have concluded from
{\sl HST} photometry for more than 1000 GC candidates
that the GC metallicity distribution is more likely bimodal in nature.}

Histograms for the goodness-of-fit statistic, ${\chi}^2_{\rm sim}$, are given in the 
right-hand panels of Figure 5. Since the absolute calibration of this statistic depends 
on the the appropriateness of our assumption of Poisson uncertainties in the observed 
and simulated distributions, it is probably best to interpret the ${\chi}^2_{\rm sim}$ 
values in a relative sense.  

\subsubsection{Dependence on LF Slope}

FBG97 have shown that, for the dozen or so gEs which are known to have
bimodal GC metallicity distributions, the location of the metal-poor peak does not
appear to obey the relation between GC metallicity and host galaxy absolute 
magnitude defined by the GCSs of fainter galaxies and the metal-rich GCSs of
gE galaxies. Instead, the metal-poor peaks for these galaxies fall in the range 
$-1.5 \lae {\rm [Fe/H]} \lae -0.5$, with little or no correlation with $M_V$. A possible 
explanation for this curious result is given in Figure 6 which shows 
the mean location of the metal-poor peak as a function of LF slope. (The 
dependence on $M^*_V$ is much weaker.) The metallicity distribution of the captured 
GCs shows a maximum in the range $-1.5 \lae {\rm [Fe/H]} \lae -0.7$,
though at fixed slope there is a scatter of $\simeq$ 0.15 dex in the peak location,
a consequence of the stochastic nature of galaxy mergers. Thus, the weak
correlation between the location of this peak and the luminosity of the
gE is a result of the fact that the mean metallicity of the metal-poor GCs is
determined primarily by: (1) the detailed merger histories of the individual gEs and; 
(2) the initial LF of the host cluster (see \S6).

To explore this issue more closely, we have measured the location of the
metal-poor GC peak as a function of both  $\alpha$ and final gE magnitude, $M_V^f$.
Figure 7 shows the data of FBG97 (open squares) along with the results
of our Monte-Carlo experiments. Although the shift in the peak 
metallicity with $M_V^f$ at fixed $\alpha$ is modest (particularly 
for steep LFs) and the intrinsic scatter in [Fe/H] is $\simeq$ 0.15 dex, 
there is a slight tendency for the brighter gEs
to capture GC populations which are more metal-rich than those accreted
by their fainter counterparts (since the brighter gEs are able to accommodate
the capture of more luminous intruder galaxies). For simplicity, we have 
assumed equal numbers of metal-rich and metal-poor GCs in the simulations; 
the observed galaxy-to-galaxy scatter in this ratio (see Table 1 of FBG97) 
is expected to weaken further the correlations shown in Figure 7, as are
differences in the initial LFs.

\section{A Possible Origin for High-$S_n$ Galaxies: The Case of NGC 4486}

Up to now, we have made no distinction between GCs acquired through mergers or through
tidal-stripping since the exact mechanism by which they are captured will have little effect
on the shape of the GC metallicity distribution (see \S2.1). However, since tidal stripping will
remove only a fraction of the GCS of an individual galaxy, the number of 
galaxies which must be stripped in order to explain the observed number of 
metal-poor GCs will always exceed the number of galaxies which must be captured.
Moreover, the GCSs of many galaxies are known to be more spatially
extended than the underlying halo light ($e.g.$, Harris 1991)
so it is possible that the capture of GCs by
tidal stripping will affect the specific frequency of the gE differently
than will capture by mergers ($i.e$, by preferentially stripping high-$S_n$ material
from the outer regions of passing galaxies).
We now explore the consequences of tidal stripping on specific frequency.

\subsection{The Initial GCSs of Low- and Intermediate-Luminosity Galaxies}

The possibility that the high specific frequencies of some centrally dominant 
galaxies can be explained through tidal-stripping of GCs from
other cluster members was first suggested by Forte, Martinez \& Muzzio (1982).
Several potential problems with this scenario were pointed out by van den Bergh (1984)
and McLaughlin, Harris \& Hanes (1994), the most serious being the objections that: 
(1) since tidal stripping will remove halo stars and GCs in equal proportion, 
such a process cannot explain why the specific frequency of NGC 4486 is roughly three 
times greater than that of other Virgo gEs; and (2) the high specific frequency
problem in NGC 4486 is not just confined to the galaxy's envelope, since even at a
galactocentric distance of $1 - 2^{\prime}$, the specific frequency is 
$S_n \simeq 10$. Given that tidal stripping has long
been recognized as providing a natural means of producing the extended
envelopes of cD galaxies ($e.g.$, Richstone 1976), it is worth
investigating this issue more closely, particularly in light of several
recent detections of probable tidally-stripped debris in galaxy clusters,
including intergalactic stars (Ferguson, Tanvir \& von Hippel 1998) and 
intergalactic planetary nebulae (Theuns \& Warren 1997; M\'endez et al. 1997; Ciardullo et al. 1998
and references therein).
The results of the planetary nebulae surveys are particularly interesting
since these studies suggest that $\sim$ 26 -- 50 \% 
of the total Virgo luminosity may reside in intergalactic stars; if
so, and if this diffuse component has a ``normal" GC specific frequency of
$S_n = 5$, then it should be accompanied by a population of 35000 -- 75000 tidally-stripped GCs.
For higher specific frequencies (see below), the total number of GCs will
of course be larger still.

Apart from this observational evidence, there are theoretical reasons to
believe that the tidal field of the cluster potential well acts to remove
matter from galaxy halos on time-scales comparable to the crossing time
(Peebles 1970; Gunn 1977; Merritt 1985). Consequently, this process, along with collisional
stripping produced during galaxy-galaxy encounters (which usually operates on 
much {\it longer} time-scales; Merritt 1984), may be responsible for the different 
kinds of tidal debris mentioned above, particularly since low- and intermediate-luminosity
galaxies may be strongly stripped by the mean cluster tidal field.

It is recognized that the GCSs of many galaxies are more extended than the 
underlying halo distributions ($e.g.$, Harris 1991). However, many of the galaxies 
upon which this conclusion is based are located in rich clusters, so it is unclear
how environment has influenced the presently-observed distribution of
GCs. In other words, are the GCSs more extended as a {\it result} of tidal stripping,
or did they {\it form} that way? Since tidal limitation by the mean
cluster field is likely to occur early on in the evolution of the cluster (Merritt 1984; 1985)
we need to consider the relative spatial extents of the halo and GC populations
of galaxies at the time of galaxy/cluster formation.

To do so, we must identify a sample of isolated galaxies having
well-studied GCSs and accurate surface brightness profiles. We therefore
follow Minniti, Meylan \& Kissler-Patig (1996) in defining a ``master dE"
galaxy using the Local Group dwarf galaxies Fornax, NGC 147, NGC 185 and 
NGC 205. To characterize the halo light in each galaxy, we adopt the 
exponential scale-lengths and central surface brightnesses given by
Caldwell et al. (1992). We then correct each galaxy to a luminosity-weighted mean scale-length
of ${\overline{\alpha_s}} = 375$ pc, and add the individual surface brightness 
profiles to create a composite profile. In converting from angular to linear 
scales, we adopt the same values for the distance and reddening of each galaxy which were used 
to derive the absolute magnitudes in \S2.1. Similarly, we scale the 
radial positions of the GCs in each galaxy in order to bring them
onto a uniform scale. The composite surface brightness profile is
shown in Figure 8, along with the corrected profiles for each galaxy
and for the entire ensemble of 24 GCs.

By considering only isolated dE galaxies ($i.e.$, systems which have probably not interacted 
strongly with nearby galaxies), we minimize concerns that their GCSs have been modified
by dynamical effects which are {\it external} to the parent 
galaxies.\altaffilmark{8}\altaffiltext{8}{The proximity of NGC 205 to M31 suggests
it may be tidally interacting with its parent galaxy. It is included 
in the definition of the master dE since in \S4.2 and 4.3 we show that past 
interactions of this sort are likely to have
preferentially removed GCs rather than field stars. Including NGC 205 in
the sample therefore leads to a conservative estimate of the initial specific frequency 
profile of the master dE galaxy.}
However, {\it internal} processes, of which orbital decay through dynamical friction
is most important, will also affect the spatial distribution of GCSs. To estimate
the importance of this effect, we use Equation 7-25 of Binney \& Tremaine (1987),
$$r{dr \over dt} = -0.428{GM \over v_{typ}} \ln{\Lambda}, \eqno{(12})$$
which gives the orbital decay time for an object moving in a circular orbit around 
a galaxy with an isothermal potential. Here $M$ is the mass of the GC in 
question, $v_{typ}$ is the typical stellar velocity
in the galaxy, and the Coulomb logarithm $\ln{\Lambda}$, is given by
$${\Lambda} = {b_{\rm max}v_{typ}^2 \over GM}.\eqno{(13})$$ For $b_{\rm max}$, which denotes
the distance at which the stellar density becomes much smaller than it is in 
the neighborhood of the orbiting GC, we adopt the effective radius of each galaxy,
given by $r_{\rm eff} = 1.678\alpha_s$ (Impey, Bothun \& Malin 1988). 
We take $v_{typ}$ to be the mean circular velocity, calculated as $\sqrt{2}\sigma_g$,
where $\sigma_g$ is the one-dimensional stellar velocity dispersion in each galaxy
(Mateo 1994).\altaffilmark{9}\altaffiltext{9}{Our adopted
value of $\sigma_{g} = 38.5$ km s$^{-1}$ for NGC 205 represents the average of
the mean ``core" and ``halo" dispersions quoted by Mateo (1994).}
Integration of the above equation gives the initial galactocentric distance, 
$R_i$, of each GC in terms of its current galactocentric distance, $R_f$,
$$R^2_i = 0.856{GM \over v_{typ}}\ln{\Lambda}~t_{\rm H} + R^2_f,\eqno{(14})$$
where $t_{\rm H}$ is the time since formation.
In practice, we calculate $R_i$ for each of the 24 GCs in the master dE using the
values of $v_{typ}$ and $b_{\rm max}$ appropriate for each dwarf, and
combine the $V$ magnitude of each GC (see Da Costa \& Mould 1988 and 
references therein)\altaffilmark{10}\altaffiltext{10} {For GCs IV, VI, VII and VIII in
NGC 185 (Ford, Jacoby \& Jenner 1978), there are no published magnitudes 
or colors. We have estimated approximate magnitudes for these objects by 
performing aperture photometry on a $V$-band image taken with the 
KPNO 4m telescope which was kindly provided by Peter Stetson.}
with our adopted distance estimates 
to derive individual masses. In doing so, we assume a constant GC mass-to-light ratio 
of $M/L_V$ = $2M_{\odot}/L_{V{\odot}}$.

Figure 9 shows the radial dependence of specific frequency in the master dE. 
Triangles indicate the net specific frequency, $S_n^f~(>R)$, outside of the marked
positions, obtained from the presently-observed galactocentric distances of the GCs.
Squares show the net specific frequency, $S_n^i~(>R)$, after 
correcting the position of each GC for the effects of dynamical friction.
Clearly, the outer regions of such intermediate-luminosity galaxies are likely to 
have consisted of ``high-$S_n$" material at the epoch of galaxy/cluster formation.
Unfortunately, the short dynamical friction time-scale for GCs originally confined to 
the inner regions of their parent galaxies
means that $S_n^i~(>R)$ is essentially unconstrained interior to $\sim$ 1 kpc.
Thus, we cannot determine if this region {\it also} exhibited high initial
specific frequencies due to the presence of GCs which have subsequently 
been destroyed.

\subsection{The Effect of Tidal-Stripping on Specific Frequency}

Merritt (1984) has considered the case of a test galaxy moving through a
cluster having a King model density profile.
In this case, tidal stripping by the mean field of the cluster
will lead to tidal radii, $r_t$, of the order
$$r_t \approx {R_c\sigma_g \over 2\sigma_c} \eqno{(15})$$
where $R_c$ is the ``core" radius of the cluster, $\sigma_c$ is its velocity
dispersion and $\sigma_g$ is the velocity dispersion of the test galaxy. For
galaxies similar to our master dE, we expect $\sigma_g \sim 30$ km s$^{-1}$,
and the {\sl ROSAT} observations of Nulsen \& B\"ohringer (1995) imply
a Virgo core radius of $R_c = 45$ kpc  for our adopted Virgo distance of 16 Mpc.
Since the velocity dispersion of early-type, and presumably virialized, galaxies 
in the vicinity of NGC 4486 is known to be $\sigma_c \simeq 600$ km s$^{-1}$
(Binggeli, Tammann \& Sandage 1987), Equation 15 predicts $r_t \sim 1$ kpc.

It is important to investigate the sensitivity of 
$r_t$ to the assumed density profile, since recent investigations into the 
structure of dark halos using high-resolution, N-body simulations have cast doubt
on the very existence of constant-density ``cores" in galaxy clusters (Navarro,
Frenk \& White 1996; hereafter NFW). These authors suggest that a model of the form
$$ \rho(r) \propto {1 \over (r/r_s)(1 + r/r_s)^2},\eqno{(16)}$$
where $r_s$ is a scale radius, is a better representation of the true dark matter 
distribution.  In Appendix A we show that the above estimate of $r_t \sim 1$ kpc 
is unchanged if we adopt a Virgo density profile of the form of Equation 16.

From Figure 9, we see that the tidally-stripped, outer regions of the master dE galaxy would have
a net specific frequency of $S_n \gae 10$, similar to that observed for NGC 4486
(McLaughlin, Harris \& Hanes 1994). Since stripping by the overall cluster 
potential will be most important near the dynamical center of the host cluster, 
it offers a natural explanation for fact that ``high-$S_n$" galaxies are 
located, without exception, near
the dynamical centers of massive clusters ($i.e.$, where the tidal forces
are greatest and where the tidal debris accumulates). Thus, it may explain the
correlations between the number of ``excess" GCs and: (1) cluster mass; and
(2) ambient mass density found by
West et al. (1995) and by Blakeslee, Tonry \& Metzger (1997).
It is worth bearing in mind, however, that the galaxies which define the 
specific frequency profile shown in Figure 9 have a median
absolute magnitude of $M_V = -15.5$, whereas the bulk of the captured
GCs associated with M87 probably originated in slightly more
luminous galaxies ($i.e.$, $M_V \sim -17$, based on Equation 3 and on the observed 
location of the metal-poor GC peak in M87). Additional
observations of the GCSs of intermediate-luminosity galaxies 
are clearly warranted.

\subsection{The Radial Profile of Tidally-Stripped GCs}

Is the spatial distribution of GCs surrounding NGC 4486 consistent with the existence 
of two chemically-distinct GC populations, one which formed along with the
body of the galaxy itself, and a second which was tidally stripped from other
cluster galaxies? In Figure 10 we show the surface density profile of NGC 4486 GCs
taken from Harris (1986). Though not as deep as
recent CCD surveys of the NGC 4486 GCS ($e.g.$, McLaughlin, Harris \& Hanes 1994;
Whitmore et al. 1995; Elson \& Santiago 1996), this study offers much greater
areal coverage and represents the best existing survey of the spatial structure of the
GCS at large distances from the galaxy's center.
Note that we have plotted {\it the total counts of unresolved sources}
in order to investigate the possibility that the GCS 
surrounding NGC 4486 has a spatial extent which exceeds that of existing wide-field surveys.
GC candidates are indicated by the open circles.
The filled squares show the arbitrarily shifted surface brightness profile for the 
galaxy itself. The dashed line represents the $R^{1/4}$ law which best fits the
halo light interior to 4\farcm5, the radius which marks the onset of NGC 4486's cD envelope. 
If we assume that the population of metal-rich GCs originally 
followed the same $R^{1/4}$ profile as the underlying halo light, we can then use the
de Vaucouleurs \& Nieto (1978) relation to define the radial profiles of both intrinsic
and captured GCs, provided we know their relative surface densities at a single 
position.\altaffilmark{11}\altaffiltext{11}{Unless the metal-rich GCs in NGC 4486 
are {\it much} more spatially extended than the underlying halo light,
this assumption will have a negligible effect on the derived radial profile of
metal-poor GCs in the outermost regions of the galaxy. For instance, if we assume
that the metal-rich GCs are {\it twice} as extended as the halo stars, the 
residual GC surface densities are reduced by $\lae$ 6\% beyond 1\farcm5.}
Such an estimate is possible using the deep {\it HST} survey of Elson \& Santiago (1996)
who obtained $VI$ photometry of 220 GCs in a field located $\simeq$ 2\farcm5
from the center of NGC 4486.\altaffilmark{12}\altaffiltext{12}
{We use the Elson \& Santiago (1996) GC photometry instead of that of Whitmore et 
al. (1995) in order to minimize concerns that the metal-rich and metal-poor GC
populations have suffered different amounts of dynamical erosion
in the innermost regions of the galaxy
($i.e.$, the Elson \& Santiago field is located $\simeq$ 2\farcm5 
from the center of NGC 4486, whereas the field of Whitmore et al. extends 
to within 14$^{\prime\prime}$ of the galaxy's center).}
According to Elson \& Santiago (1996), roughly 44\% of the GCs in this field belong 
to the metal-rich component. 
Since the Harris (1986) star counts imply a {\it total} GC surface density of 26 arcmin$^{-2}$
at $R$ = 2\farcm5, we take the surface density profile of
metal-rich GCs in NGC 4486 to be $\log{\Sigma} = 1.452 - 3.33[(R/1\farcm6)^{1/4} - 1]$.
This profile is shown by the dotted line in Figure 10. 
The residuals between the measured GC profile of Harris (1986) --- which is composed of both metal-poor
and metal-rich clusters --- and that expected for the metal-rich GCs are
plotted as filled triangles. 

How does the radial profile of these ``residual" GCs compare to that of the dark
matter in the Virgo cluster core?
To answer this question, we make use of the dynamical study of Cohen \& Ryzhov (1997)
who used radial velocities for more than 200 GCs to measure the cumulative mass
distribution within $\sim$ 40 kpc of NGC 4486. A least-squares fit to the data in
their Table 8 yields a relation of the form
$$\log{M(r)} = {a} + {b}\log{r}\eqno{(17})$$
for the total mass, $M(r)$, within a de-projected radius $r$. Assuming a spherically
symmetric mass distribution, this implies a density profile, $\rho(r)$, of the form
(cf. Equation 16)
$$\rho(r) = {ab \over 4\pi}r^{b-3}\eqno{(18})$$
where $a = 11.08\pm0.08$ and $b = 1.70\pm0.15$ (1$\sigma$ uncertainties). The surface
density, $\Sigma(R)$, of a population of objects which follows the above mass
distribution is therefore
$$\Sigma(R) \propto R^{-0.30\pm0.15}.\eqno{(19})$$ 
In Figure 10 we plot this radial profile as the solid curve. The associated 1$\sigma$ 
confidence limits are indicated by the upper and lower dotted curves.

Despite the fact that the uncertainties in the derived mass 
distribution are rather large, and that the radial profile of the ``residual" GCs was 
calculated under the assumption of negligible backround contamination, the agreement
beween the two profiles is striking.
Unfortunately, the photographic survey of Harris (1986) was carried out in a single bandpass, 
so no information is available on the colors of the GCs at large radii. Clearly, 
it is important to obtain deeper, multi-color CCD images of the Virgo cluster
core having a comparable, or greater, field size. Such observations are now
feasible using wide-field mosaic CCD cameras which are available on
many 4m-class telescopes, and would provide a straightforward test of the prediction
that appreciable numbers of metal-poor, tidally-stripped GCs reside at large distances 
from NGC 4486. For instance, at a distance of $R$ =  40$^{\prime}$ we expect a surface
density of $\sim$ 5 GCs arcmin$^{-2}$. At such distances, captured GCs are expected to
greatly outnumber intrinsic GCs, so that the metallicity distribution of the outermost GCs
is predicted to be metal-poor and essentially unimodal.

Conversely, the intrinsic population of GCs should dominate near the very center of
the galaxy. Based on Figure 10 and on the GC surface densities for the inner regions 
of NGC 4486 reported by McLaughlin (1995), we estimate that the metal-rich GCs should 
outnumber the metal-poor ones inside $R \sim$ 30$^{\prime\prime}$, with the exact ratio 
depending on the exponent governing the dependence of mass density on radius.
Note that only those GCs 
initially within $R \sim 0\farcm1$ are expected to have had their
orbits decay to the galaxy center through dynamical friction (Equation 14; 
Lauer \& Kormendy 1986). 
It is also worth pointing out that Equation 19 predicts a surface density for the metal-poor 
GCs of 15$^{+7}_{-4}$ arcmin$^{-2}$ at a galactocentric distance of $R = 1\farcm5$; since 
the total GC surface density at this radius is approximately 45 GCs arcmin$^{-2}$, the
implied surface density of intrinsic GCs is $\simeq$ 30 arcmin$^{-2}$.
Thus, the measured specific frequency of $S_n = 10\pm1$ at this distance (McLaughlin, 
Harris \& Hanes 1994) would correspond to $S_n = 6.5^{+1.2}_{-0.9}$ for the intrinsic 
GCS, which is more in accord with that found for other Virgo gE galaxies.

\section{A Comparison to Merger-Induced GC Formation Models}

There have been several recent reviews of models for the formation 
of gEs and their associated GCSs 
($e.g.$, Harris, Pritchet \& McClure 1995; FBG97; 
Blakeslee, Tonry \& Metzger 1997). One model which has received particularly
close attention  --- not only in the context of the so-called ``high-$S_n$ problem" but
also in terms of bimodal GC metallicity distributions --- is that of Ashman \& Zepf (1992)
which contends that numerous GCs {\it are formed} during mergers of gas-rich galaxies.
The model proposed here also involves mergers, but makes very different 
predictions for the formation and evolution of gE galaxies and their
systems of GCs. Therefore, we include below a brief comparison of the present model and
that of Ashman \& Zepf (1992), focusing on the well-studied pair of
Virgo gE galaxies NGC 4472 and NGC 4486.

Perhaps the strongest evidence against the notion that the formation of new GCs in
mergers can explain both the bimodal GC metallicity distributions and extraordinarily
large GCSs of some gE galaxies
is that {\it both} NGC 4472 and
NGC 4486 exhibit strongly bimodal GC metallicity distributions despite the fact that
NGC 4486, though $\sim$ 20\% less luminous than NGC 4472, has more than twice as many
GCs. If repeated mergers gave rise to the metallicity distributions observed in both galaxies, 
then how did NGC 4486 manage to form GCs almost three time more efficiently than its
Virgo counterpart? Indeed, the dichotomy is even more pronounced than such arguments 
suggest since the merger rate in the vicinity of NGC 4486 is likely to have been 
negligibly small once the deep potential well of the cluster formed 
($i.e.$, the merger time-scale depends very strongly the cluster velocity
dispersion; see Equation 31 of Merritt 1985). 
In the model proposed here, the proximity of NGC 4486 to the dynamical center of Virgo 
offers a straightforward explanation for the different number of GCs associated with
the two galaxies.

For gEs having specific frequencies greater than $\sim {\overline{S_n}}$, we predict 
that the ``excess" GCs should be metal-poor. This prediction is contrary to that of 
Ashman \& Zepf (1992) who argue that the excess GCs should be metal-rich, having 
formed from chemically-enriched gas. Figure 3 of FBG97 demonstrates that, among the
dozen or so gE galaxies which are known to have bimodal GC metallicity distributions,
the metal-poor GCs dominate the overall GC population in high-$S_n$ galaxies.
In addition, an age-metallicity relation is expected for GCs in the Ashman \& Zepf 
(1992) model, in the sense that the metal-rich GCs should be younger than their
metal-poor counterparts. Such a trend is not observed for the NGC 4486 GCS (Cohen,
Blakeslee \& Ryzhov 1998), nor is one expected within the framework of the
model proposed here.

\section{Additional Predictions}

In addition to our predictions that there should exist appreciable numbers of
metal-poor GCs at large distances from NGC 4486, and that the ratio 
of metal-poor to metal-rich GCs should increase steadily with distance from
the galaxy's center (see \S4.3), there are a number of other observations which 
may be used to test our model for the origin of the metal-poor GCs 
associated with gE galaxies.
First, we expect no dwarf ellipticals to show bimodal GC metallicity
distributions since these galaxies are unlikely to have complex merger histories.
In addition, since the metal-rich GCs associated with gE galaxies are assumed to 
have formed {\it in situ}, during a ``standard" dissipational collapse scenario, 
they may show a radial gradient in metallicity and some net rotation, {\it depending 
on the amount of dissipation in the collapse of the proto-galactic gas cloud} 
(see, $e.g.$, Majewski 1993). Based on Figure 6, we also predict a correlation
between the location of the metal-poor GC peak and the slope of the initial
galaxy LF, in the sense that steeper LFs produce lower metallicities for the 
captured GCs. However, testing this claim will require large sample sizes
since the expected trend will be diluted by differences in the merger 
histories of individual galaxies. In a companion paper (Marzke, C\^ot\'e \& West 1998),
we discuss the formation of cD galaxies by tidal stripping and mergers, and show 
that a similar trend should exist between initial LF slope and cD envelope color.

The metal-poor GCs surrounding NGC 4486 are predicted to have been tidally
stripped from other cluster galaxies by the overall Virgo cluster potential.
Figure 10 suggests that their spatial distribution is consistent with the
idea that they are tidal debris, and that they follow the same surface
density profile as does the mass in the Virgo core. Thus, the metal-poor 
GCs in NGC 4486 are expected to show a larger velocity dispersion than the metal-rich GCs.
Radial velocities for the most distant GCs ($i.e.$, those beyond $R \sim 25^{\prime}$)
should have a dispersion comparable to those for the Virgo cluster galaxies.
Recently, Kissler-Patig (1997) has presented evidence for such an effect
in the Fornax cluster: the distant, and predominantly metal-poor, 
GCs surrounding the cD galaxy NGC 1399 have a velocity dispersion
which is indistinguishable from that of the innermost Fornax galaxies.

Finally, since tidal stripping of GCs is dominated by the overall cluster 
potential and not by that of the gE galaxy itself, tidally stripped 
GCs should accumulate in the potential well of {\it any} massive cluster, regardless
of the whether or not the potential well coincides with the location of a bright
cluster galaxy (Merritt 1984; White 1987; West et al. 1995).

\section{Summary}

Many, perhaps most, gE galaxies have GCSs which show bimodal metallicity 
distributions.  We argue that such bimodal distribution may arise from
the capture of other GCs, either through mergers or 
tidal stripping. For reasonable choices for the LF of galaxies in the host cluster 
and for the dependence of GC metallicity on parent galaxy luminosity, such
captured GCs define a metal-poor population which has a maximum in the range
$-1.5 \lae {\rm [Fe/H]} \lae -0.7$. 
Since the precise location of this maximum depends 
primarily on the LF of the surrounding galaxy cluster, mergers and tidal stripping provide a 
natural explanation for not only the origin of the metal-poor peak, but also its 
poor correlation with the luminosity of the gE (FBG97). {\it Our principal conclusion is 
that it is possible to explain bimodal GC metallicity distributions without resorting to 
the formation of new GCs in mergers or by invoking multiple bursts of GC formation.}

A comparison between the simulated and the observed GC metallicity distributions 
for the well-studied galaxy NGC 4472 shows excellent agreement, particularly for steeper LFs. 
We argue that the weaker cluster tidal field near NGC 4472 compared
to NGC 4486 is responsible for the disparity in the sizes of the GCSs of these
two galaxies since tidal stripping may, in some cases, lead to substantial 
increases in specific frequency. This is particularly true if the metal-poor 
GCs in gEs were captured through tidal stripping
of low- and intermediate-luminosity galaxies since the GCSs of such galaxies are 
likely to have been more extended than their constituent stars at the epoch of 
formation.  It would seem the time is ripe to re-examine the dynamical evolution of 
GCSs using improved numerical simulations ($i.e.$,
existing N-body experiments suffer from limited resolution and restricted dynamic range; 
Muzzio 1987). With larger simulations and more sophisticated algorithms
it will be possible to investigate the time evolution of the size, spatial 
extent and metallicity distribution of the GCSs associated with bright cluster ellipticals.

We describe a new method for placing limits on the fraction of luminous matter which 
has been captured by individual gE galaxies. An application of the method to 
NGC 4472 suggests that as much as $\sim$ 65\% of its present-day luminosity could 
have been acquired in this way.
This technique may provide a useful tool for studying the merger histories of gE galaxies.

\acknowledgments
 
The authors thank John Blakeslee, Greg Bothun, Doug Johnstone, Jim Hesser, 
Dean McLaughlin, David Merritt, Sterl Phinney, Hans-Walter Rix, Scott Tremaine,
Sidney van den Bergh and an anonymous referee for helpful comments which improved the paper.
P.C. gratefully acknowledges support provided by the Sherman M. Fairchild Foundation.
Additional support for this work was provided to R.O.M. by NASA through grant No. 
HF-0.096.01-97A from the Space Telescope Science Institute, which is operated by the 
Association of Universities for Research in Astronomy, Inc., under NASA contract NAS5-26555.
MJ.W. acknowledges financial support from the Natural Sciences and Engineering
Research Council of Canada and the Canadian Institute for Theoretical Astrophysics.

\appendix
\section{Tidal Radii in an NFW Potential}
NFW have suggested that the dark matter distribution in galaxy clusters is given by 
$$ \rho(r) = {\rho_{\rm crit}\delta_c \over (r/r_s)(1 + r/r_s)^2}\eqno{({\rm A-1})}$$
where ${\rho}_{\rm crit}$ is the critical density, $\delta_c$ is the
dark halo's characteristic overdensity and $r_s$ its scale radius.
 
The tidal radius, $r_t$, of a galaxy moving in a cluster potential $\phi(r)$ is
given by (Merritt 1984)
$$r_t = {{\alpha}{\beta}{\sigma_g} \over \sqrt{2}} \Big [
{3 \over r}{d{\phi(r)} \over dr} - 4{\pi}G{\rho(r)} \Big ]^{-1/2} \eqno{({\rm A-2})}$$
where $\alpha$ and $\beta$ are parameters related to the internal mass distribution
of the test galaxy. Using Poisson's equation,
$${1 \over r^2}{d \over dr} \Big \lbrace r^2 {d{\phi(r)} \over dr} \Big \rbrace =
4{\pi}G{\rho}(r), \eqno{({\rm A-3}})$$
it is possible to calculate the potential gradient
$${d{\phi} \over dr} = {4{\pi}G{\rho_{\rm crit}}{\delta_c}{r_s^3} \over r^2}
\Big [ {\rm ln}(1 + r/r_s) + {1 \over (1 + r/r_s)} - 1 \Big ] \eqno{({\rm A-4}})$$
corresponding to this density profile.
Thus, Equation A-2 implies a tidal radius of
$$r_t = {{\alpha}{\beta}{\sigma_g} \over H_0{\sqrt{3{\delta_c}}}}
\Big [ {3r_s^3 \over r^3} \Big ( {\rm ln}(1 + r/r_s) + {1 \over (1 + r/r_s)} - 1 \Big )
- {1 \over (r/r_s)(1 + r/r_s)^2}\Big ]^{-1/2} \eqno{({\rm A-5})}$$
where we have made use of the fact that $\rho_{\rm crit} = 3H_0^2/8{\pi}G$.
 
The central velocity dispersion of $\simeq$ 600 km s$^{-1}$ for early-type galaxies
in the Virgo cluster corresponds to a halo circular velocity of $\simeq$ 850
km s$^{-1}$. This most closely resembles model 16 of NFW,
which has a scale radius of $r_s = 190$ kpc and a characteristic
overdensity of $\delta_c = 28200$. 
Thus, the tidal radius in the vicinity of NGC 4486's cD envelope is found to be 
$r_t \sim 1$ kpc, assuming $H_0 = 75$ km s$^{-1}$ Mpc$^{-1}$, $\sigma_g = 30$
km s$^{-1}$ and $\alpha \simeq \beta \simeq 1$.
The agreement with that found from Equation 15 is not surprising since 
NFW have shown that the density profile
given by Equation A-1 differs significantly from isothermal only
at large and small radii.

\begin{deluxetable}{crrrrrrr}
\footnotesize
\tablecaption{Dependence of GC Metallicity on Parent Galaxy Luminosity \label{tbl-1}}
\tablewidth{0pt}
\tablehead{
\colhead{Galaxy} &  \colhead{Type}   & \colhead{$M_V^i$} & 
\colhead{N$_{\rm gc}$}  & \colhead{${\overline{{\rm [Fe/H]}}}$} & \colhead{${\sigma}_i$} &
\colhead{Ref ($M_V^i$)\tablenotemark{a}} & 
\colhead{Ref(${\rm [Fe/H]}$)\tablenotemark{a}}
} 
\startdata
{\it Dwarf Ellipticals}~~~~~~   &   &             &     &                  &                &      &     \nl
Fornax      & dE0     & -12.7$\pm$0.4   &   5 &  -1.77$\pm$0.12  &  0.20$\pm$0.07 & 17   & 1,7 \nl
Sagittarius & dE      & -13.2$\pm$0.5   &   3 &  -1.79$\pm$0.13{\tablenotemark{b}}&  0.21$\pm$0.09 & 13,18 & 5  \nl
M81-F8D1        & dE      & -14.25$\pm$0.26 &   1 &  -1.8$\pm$0.3    &                & 2    & 2  \nl
NGC 147      & E5p     & -15.45$\pm$0.23 &   2 &  -2.06$\pm$0.33  &  0.27$\pm$0.16 & 6,12 & 4  \nl
NGC 185      & E3p     & -15.47$\pm$0.23 &   5 &  -1.61$\pm$0.23  &  0.31$\pm$0.11 & 6,16 & 4  \nl
VCC 1254     & dE0,N   & -16.65$\pm$0.20 &  13 &  -1.51$\pm$0.13  &  0.45$\pm$0.09 & 9,10 & 9  \nl
NGC 205      & E5p     & -16.61$\pm$0.32 &   6 &  -1.48$\pm$0.09  &  0.14$\pm$0.06 & 6,15 & 6  \nl
VCC 1386     & dE3,N   & -16.99$\pm$0.21 &   8 &  -1.43$\pm$0.10  &  0.21$\pm$0.52 & 9,10 & 9  \nl
NGC 3115 DW1    & dE1,N   & -17.62$\pm$0.14 &  24 &  -1.25$\pm$0.13  &  0.65$\pm$0.09 & 3,8  & 8  \nl
{\it Lenticulars}~~~~~~~~~~~~~   &        &                 &     &                  &                &      &     \nl
NGC 1380        & S0 & -21.0$\pm$0.3     &     &  0.15$\pm$0.5    &  &  14  & 14 \nl
{\it Giant Ellipticals}~~~~~~~   &         &                 &     &                  &                &      &     \nl
NGC 1052     & E4      & -20.25$\pm$0.3  &     &  -0.4$\pm$0.2      &                  &  11   &  11  \nl
NGC 1399     & E1p     & -20.85$\pm$0.3  &     & -0.57$\pm$0.2{\tablenotemark{c}}      &                  &  11   &  11  \nl
NGC 1404     & E1      & -20.1$\pm$0.3   &     & -0.2$\pm$0.2       &                  &  11   &  11  \nl
NGC 3311     & E+2     & -21.4$\pm$0.3   &     &  0.15$\pm$0.2      &                  &  11   &  11  \nl
NGC 3923     & E4-5    & -21.4$\pm$0.3   &     &  0.0$\pm$0.2       &                  &  11   &  11  \nl
NGC 4472     & E2      & -21.6$\pm$0.3   & 1774& -0.05$\pm$0.2      &  0.32$\pm$0.10{\tablenotemark{d}}   &  11   &  11  \nl
NGC 4486     & E+0-1p  & -21.8$\pm$0.3   &     &  0.0$\pm$0.2{\tablenotemark{e}}      &                   &  11   &  11  \nl
NGC 4494     & E1-2    & -20.0$\pm$0.3   &     & -0.3$\pm$0.2       &                  &  11   &  11  \nl
NGC 5128     & E0p     & -21.0$\pm$0.3   &     & -0.1$\pm$0.2       &                  &  11   &  11  \nl
NGC 5846     & E0-1    & -22.3$\pm$0.3   &     & -0.2$\pm$0.2       &                  &  11   &  11  \nl
IC 1459      & E3      & -20.5$\pm$0.3   &     &  0.0$\pm$0.2       &                  &  11   &  11  \nl
\enddata
 
 
\tablenotetext{a}{References for Table 1: 
(1) Buonanno et al. 1985;
(2) Caldwell et al. 1998;
(3) Ciardullo, Jacoby \& Tonry 1993;
(4) Da Costa \& Mould 1988;
(5) Da Costa \& Armandroff 1995;
(6) de Vaucouleurs et al. 1991;
(7) Dubath, Meylan \& Mayor 1992;
(8) Durrell et al. 1996;
(9) Durrell et al. 1997;
(10) Ferrarese et al. 1996;
(11) Forbes, Brodie \& Grillmair 1997;
(12) Han et al. 1997;
(13) Ibata, Gilmore \& Irwin 1995;
(14) Kissler-Patig et al. 1997;
(15) Lee 1996;
(16) Lee, Freedman \& Madore 1993;
(17) Mateo et al. 1991;
(18) Mateo et al. 1996;
}
\tablenotetext{b}{Following Durrell et al. (1997) we omit Terzan 7 from the 
calculation of ${\overline{{\rm [Fe/H]}}}$ since its high metallicity
(Da Costa \& Armandroff 1995), younger age (Layden \& Sarajedini 1997) and 
marginally discrepant velocity (Da Costa \& Armandroff 1995) suggest that it may 
not be a true Sagittarius GC.}
\tablenotetext{c}{Shows marginal evidence for a trimodal GC metallicity distribution 
(Forbes, Brodie \& Grillmair 1997). We combine their intermediate- and metal-rich 
populations and find that 55\% of the GCs comprise a metal-rich component with peak at  [Fe/H] = 
$-0.57$.} 
\tablenotetext{d}{We adopt the dispersion of the metal-rich component of the 
double Gaussian which best fits the photometrically-derived GC metallicity
distribution of Geisler, Lee \& Kim (1996).}
\tablenotetext{e}{Shows marginal evidence for a trimodal GC metallicity distribution (Forbes, 
Brodie \& Grillmair 1997). We fit a double Gaussian to the {\sl HST} data 
of Whitmore et al. (1995) and adopt [Fe/H] = 0.0 as the location of the metal-rich peak; 53\%
of the GCs in this sample belong to the metal-rich component.}
 
\end{deluxetable}
 
\clearpage

\clearpage

\clearpage
 
\plotone{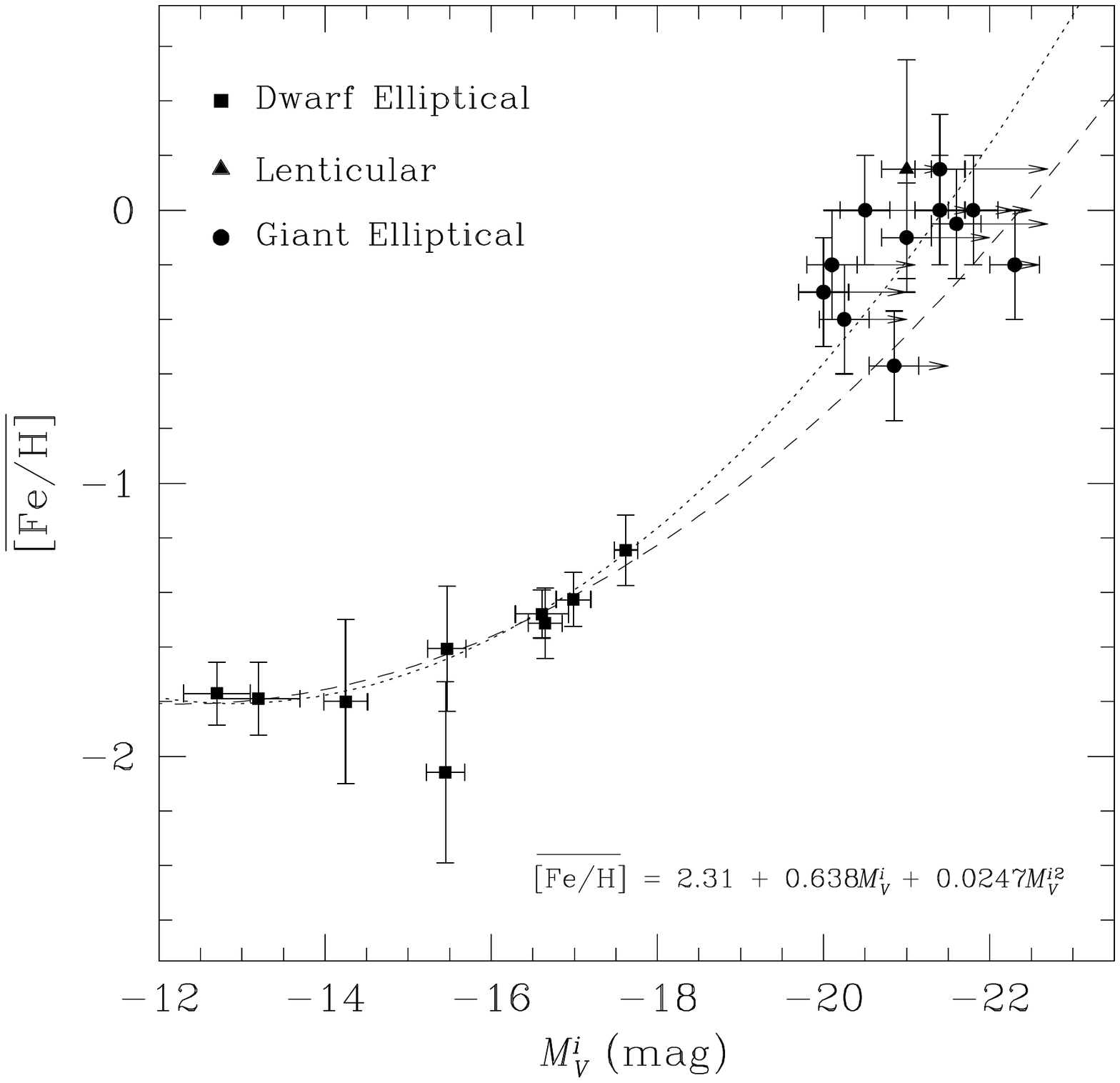}

\figcaption[cmw01.eps]{
Mean GC metallicity, ${\overline{{\rm [Fe/H]}}}$, plotted against absolute visual 
magnitude, $M_V^i$, of the parent galaxy. For the gE and lenticular galaxies, we plot 
the mean metallicities of their metal-rich GCs against the absolute visual 
magnitudes given by Equation 2. The dotted line shows the second-order polynomial 
which best fits the data, taking into account the uncertainties in both observables. 
The uncorrected magnitudes for the giant galaxies are indicated by the arrows (see text for details).
Some of the scatter among the corrected magnitudes for the giant galaxies
undoubtedly reflects the presence of radial gradients in the ratio of metal-rich 
and metal-poor GCs ($i.e.$, the GCSs are not observed over the same metric 
radii in all of the galaxies).
The dashed line shows the ${\overline{{\rm [Fe/H]}}}$-$M_V^i$ relation 
obtained using the uncorrected magnitudes.
\label{fig1}}

 
\plotone{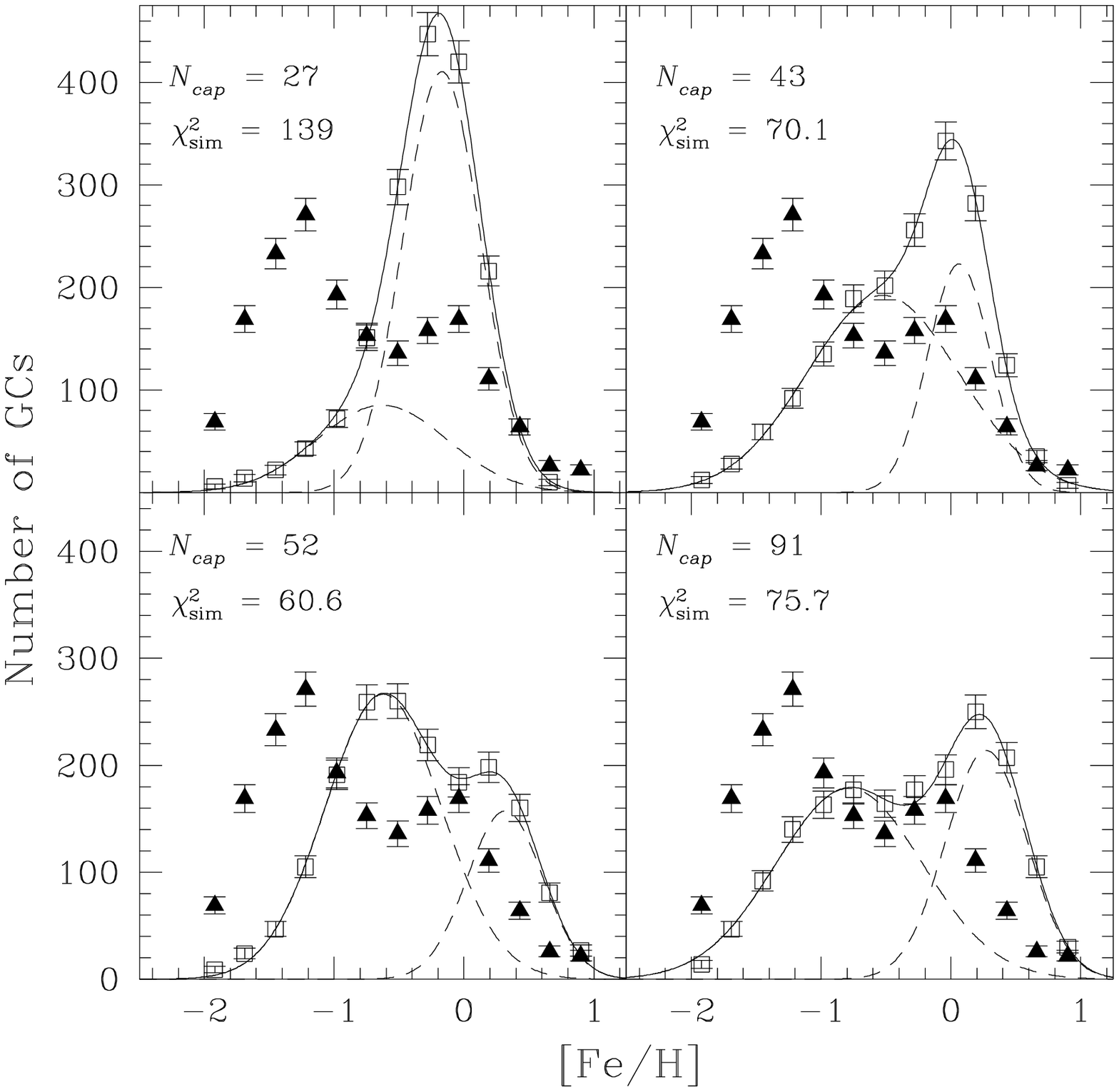}
 
\figcaption[cmw02.eps]{
Observed and simulated GC metallicity distributions for NGC 4472. The 
filled triangles indicate the observations of Geisler, Lee \& Kim (1996) 
while the open squares indicate simulated GC metallicity distributions for 
an assumed LF slope of $\alpha = -1.2$. The simulated data have been binned in
the same manner as the real data. The simulated GC metallicity distributions shown in
the four different panels are a representative sample for this choice of LF slope.
The best-fit double Gaussian is shown as the solid line, while the dotted lines 
indicate the two separate components. The total number of captured galaxies, 
$N_{\rm cap}$, is recorded in the upper left corner of each panel, along with the 
goodness-of-fit statistic $\chi^2_{\rm sim}$ (see text for details).
\label{fig2}}

 
\plotone{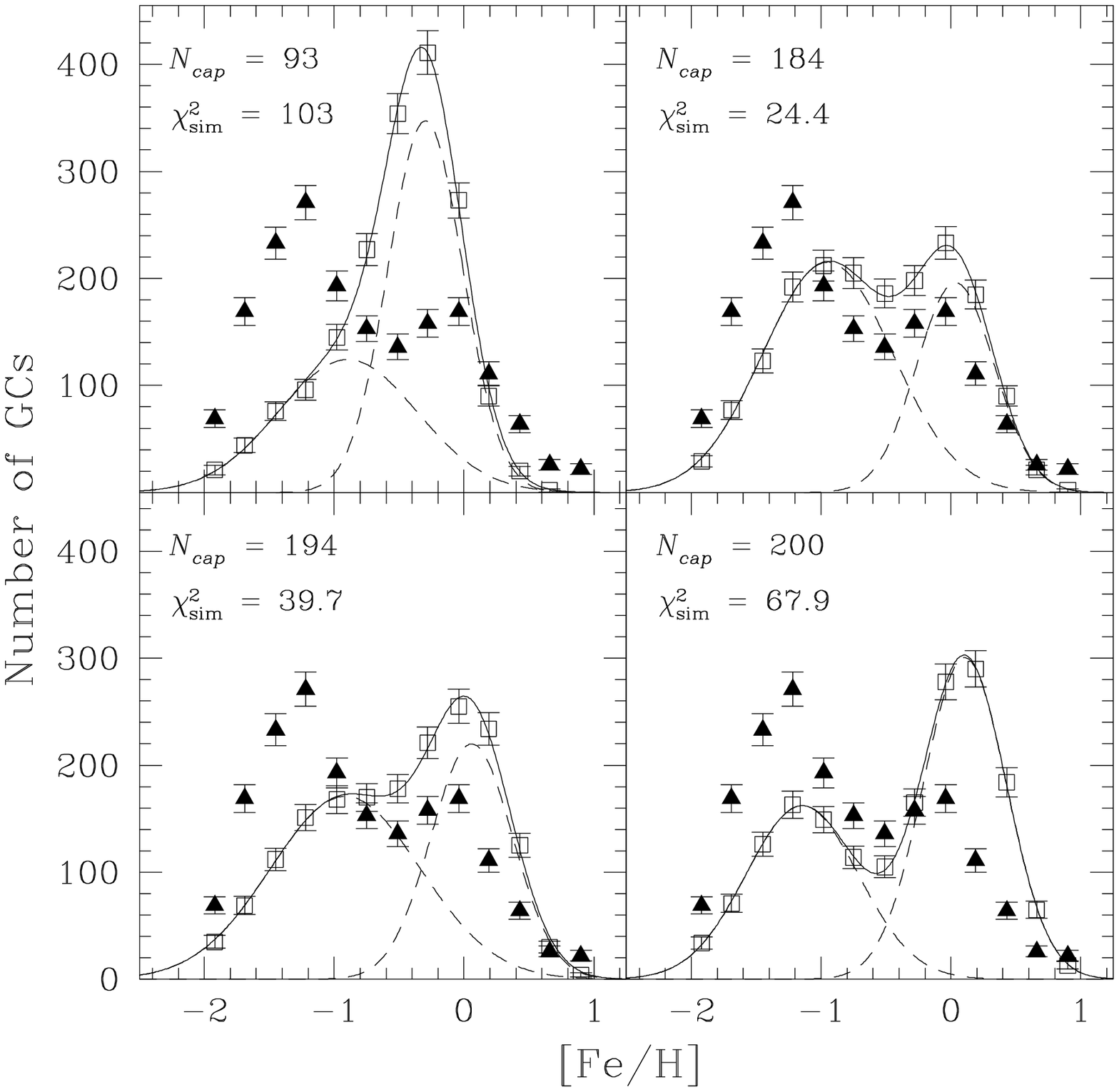}
 
\figcaption[cmw03.eps]{
Same as in Figure 2 except for an LF slope of $\alpha = -1.5$.
\label{fig3}}

 
\plotone{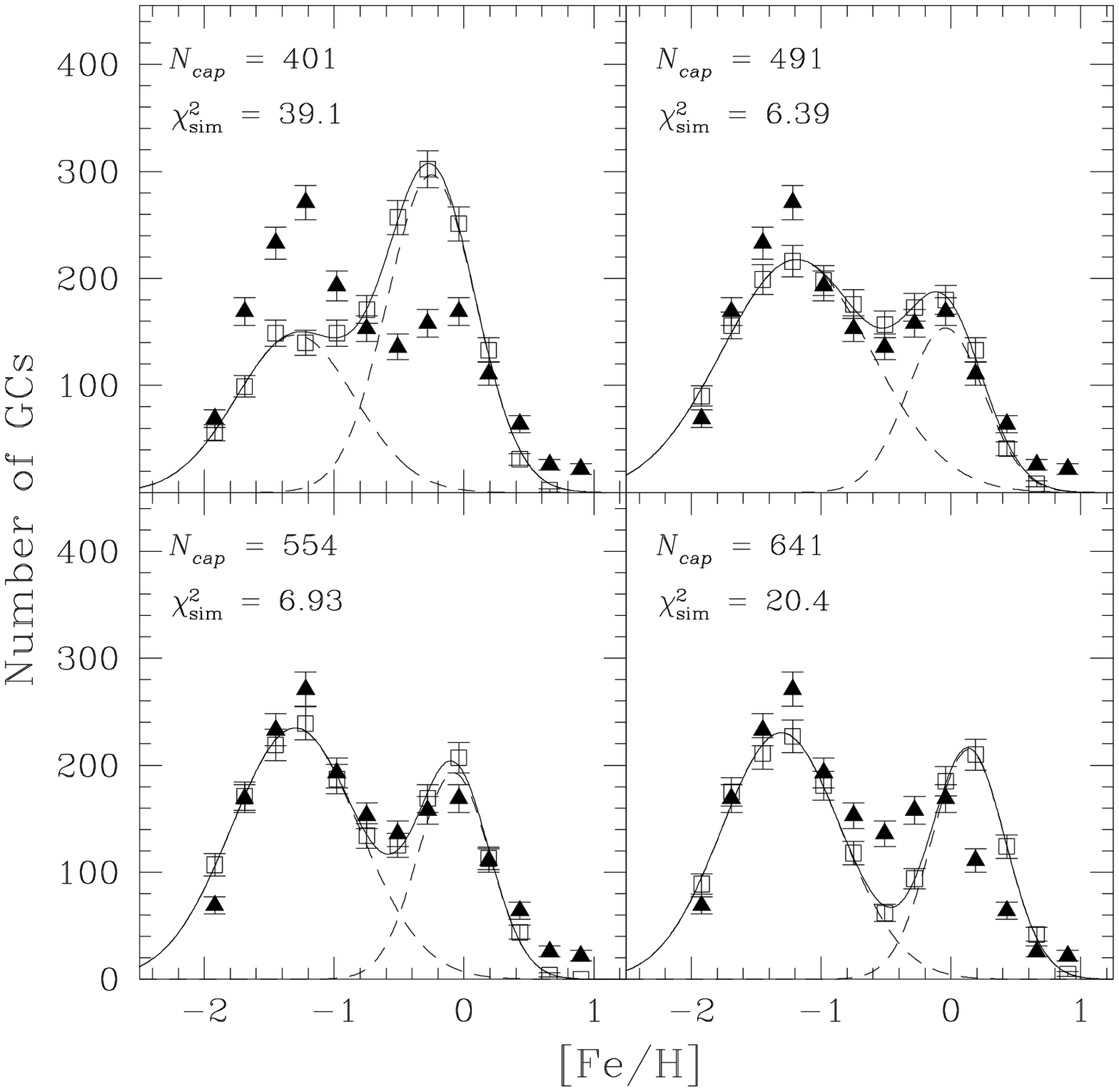}
 
\figcaption[cmw04.eps]{
Same as in Figure 2 except for an LF slope of $\alpha = -1.8$.
\label{fig4}}

 
\plotone{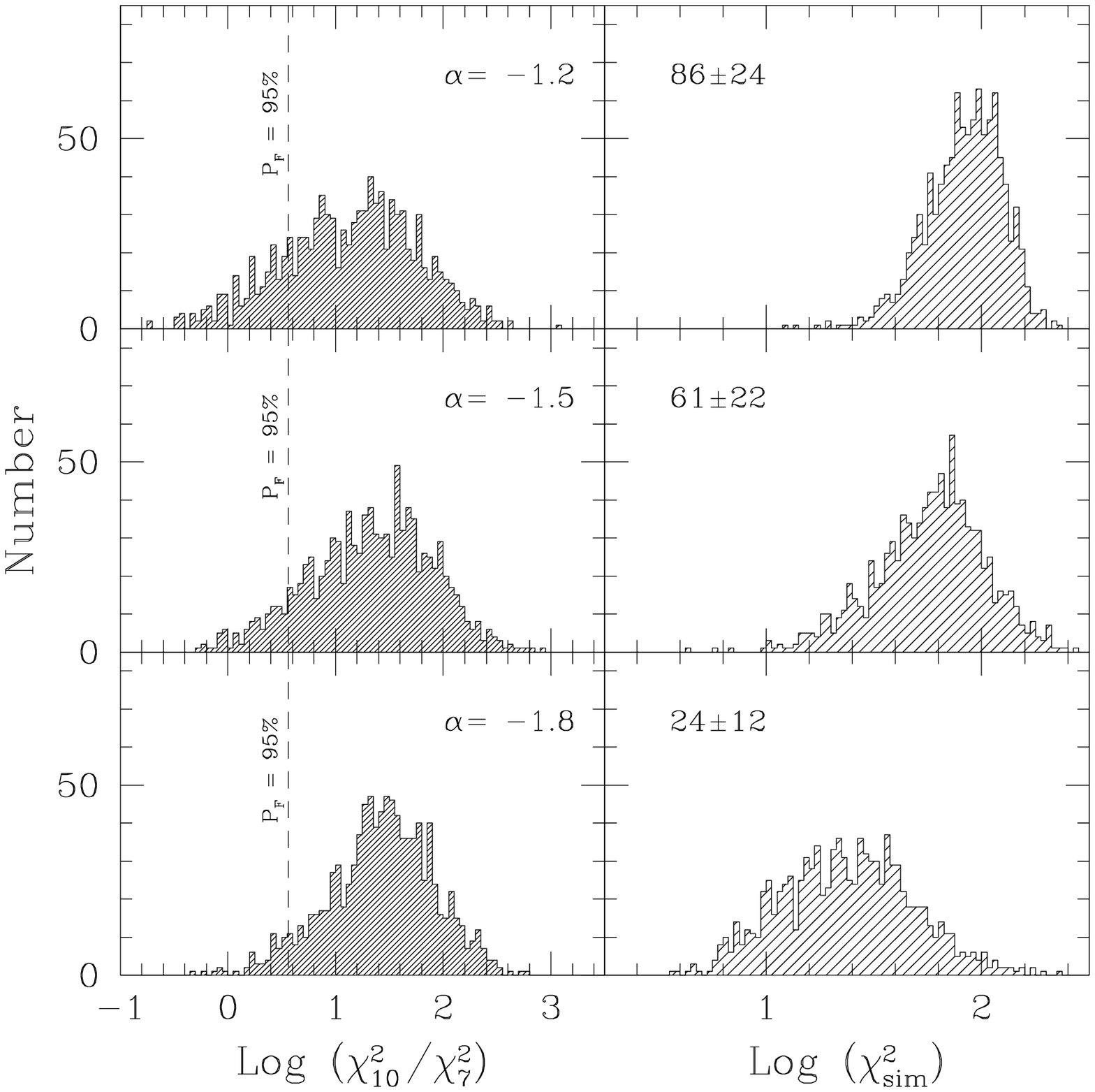}
 
\figcaption[cmw05.eps]{
(Left Panels) Histograms of the logarithm of the ratio $\chi^2_{10} / \chi^2_7$ for
1000 simulated metallicity distributions. For each simulation, the reduced chi-squared of the best-fit
single Gaussian, $\chi^2_{10}$, is measured along with the reduced chi-squared of
the best-fit double Gaussian, $\chi^2_{7}$.
The three different panels correspond to input LF slopes of $\alpha = -1.2, -1.5$ and $-1.8$.
The dashed line in each panel indicates $\chi^2_{10} / \chi^2_7 = 3.64$; 
for ratios greater than this, the inclusion of the three additional free parameters
is justified with 95\% confidence according to an $F$-ratio test.
(Right Panels) 
Histograms of the logarithm of the goodness-of-fit statistic, $\chi^2_{\rm sim}$, for the observed
and simulated distributions for NGC 4472. In each case, the LF slope
is indicated in the panel to its left.
The median value of $\chi^2_{\rm sim}$ is given in the upper left corner of each panel, along with half the
difference between the upper and lower quartiles.
\label{fig5}}

 
\plotone{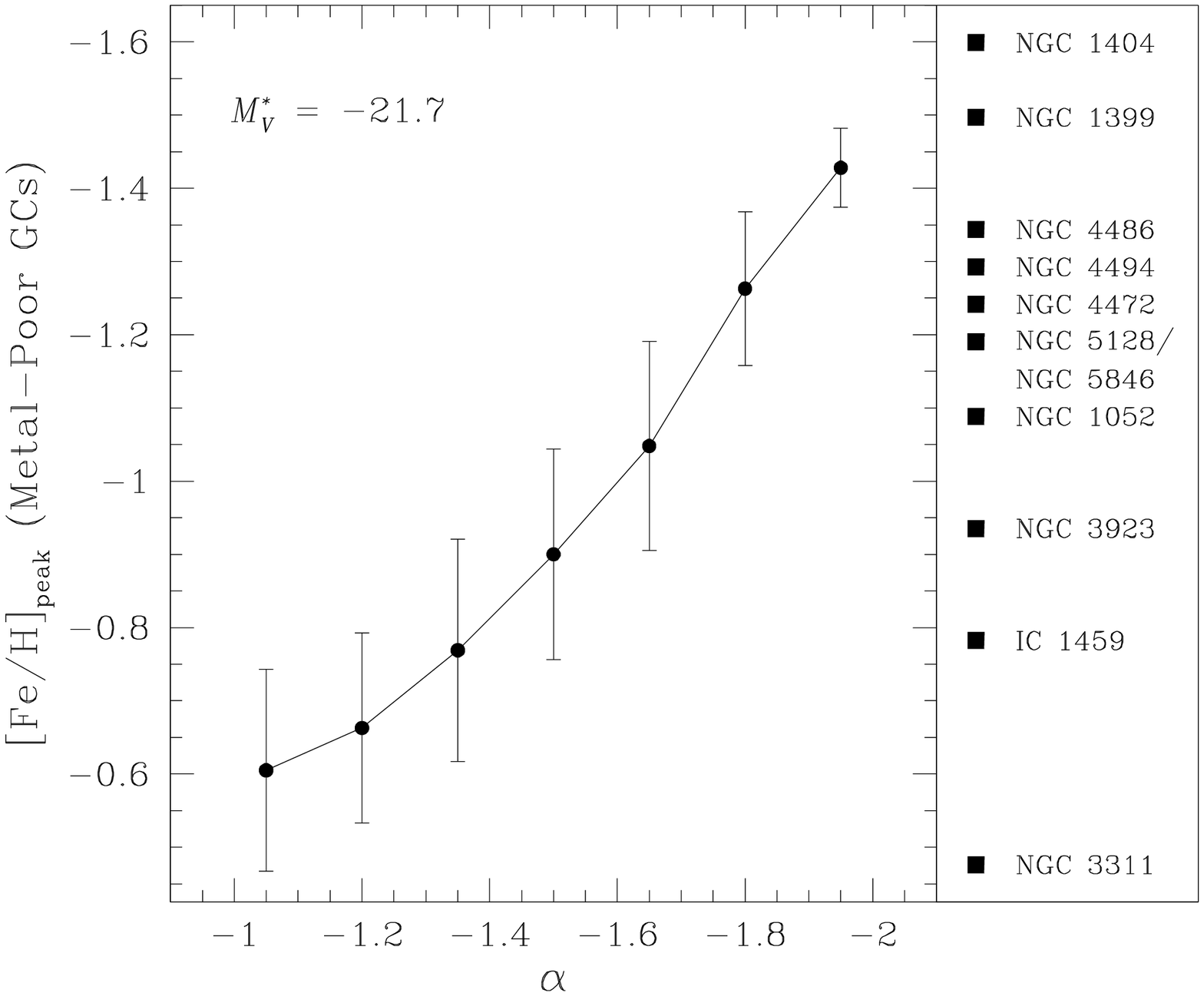}
 
\figcaption[cmw06.eps]{
Variation in the median metallicity of captured GCs as a function of LF slope
for a gE galaxy similar to NGC 4472.
The errorbars shows the upper and lower quartiles at each point.
The filled squares in the right panel show the location of the metal-poor 
peak for the gE galaxies listed in Table 1 of Forbes, Brodie \& Grillmair (1997). 
\label{fig6}}

 
\plotone{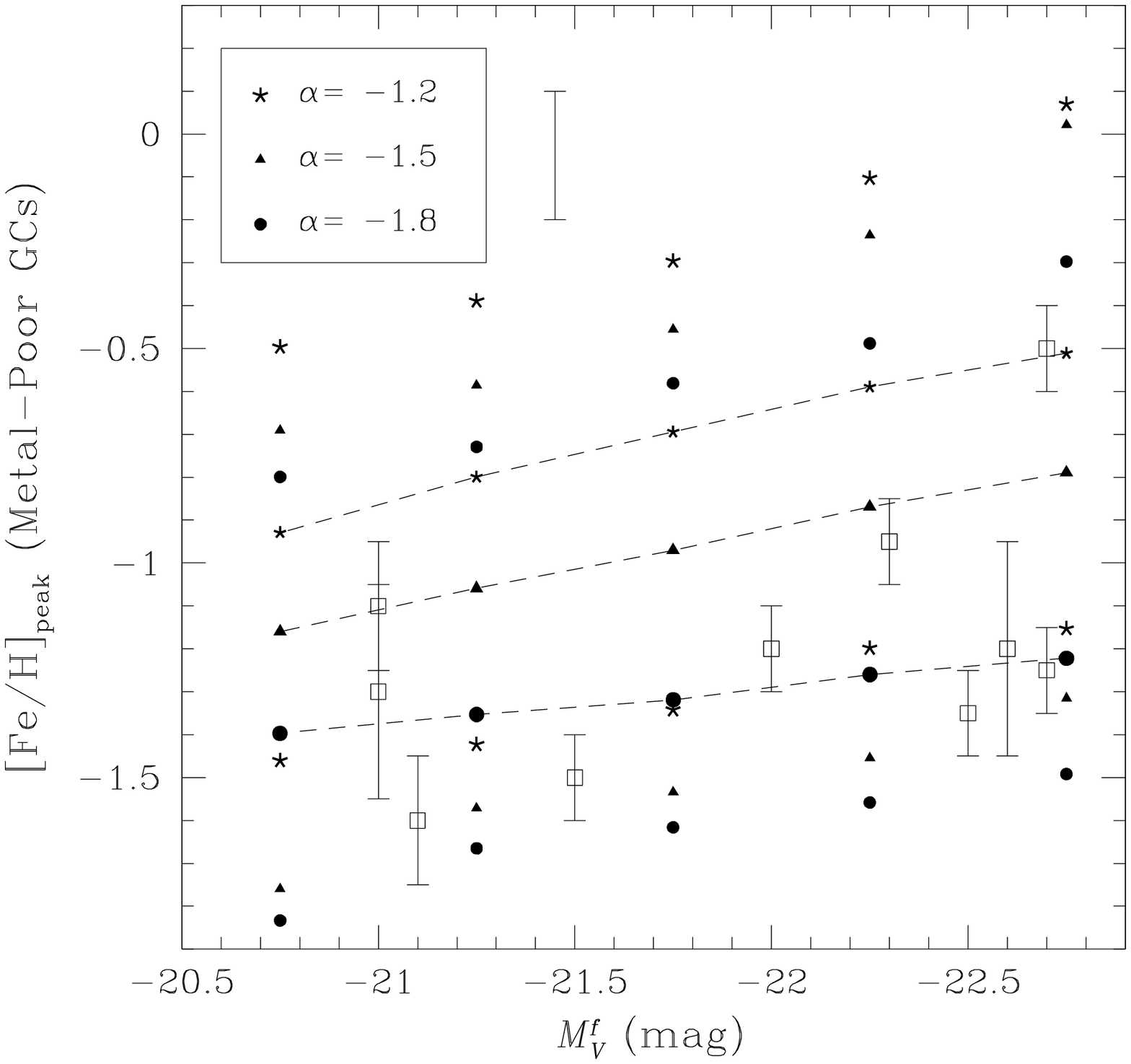}
 
\figcaption[cmw07.eps]{
Location of the metal-poor peak as a function of the final absolute magnitude, $M_V^f$, 
of the gE galaxy. One thousand simulated datasets were generated at each $\alpha$ and $M_V^f$, 
assuming equal numbers of metal-rich and metal-poor GCs. The dashed lines indicate the 
median peak location for LF slopes of $\alpha = -1.2, -1.5$ and $-1.8$. 
The errorbar at the top of the figure shows the typical standard deviation at each point.
The upper and lower points at each $M_V^f$ indicate the 99\% confidence limits
according to the simulations.  The open squares show the metal-poor peaks 
for the gE galaxies listed in Table 1 of Forbes, Brodie \& Grillmair (1997).
\label{fig7}}


\plotone{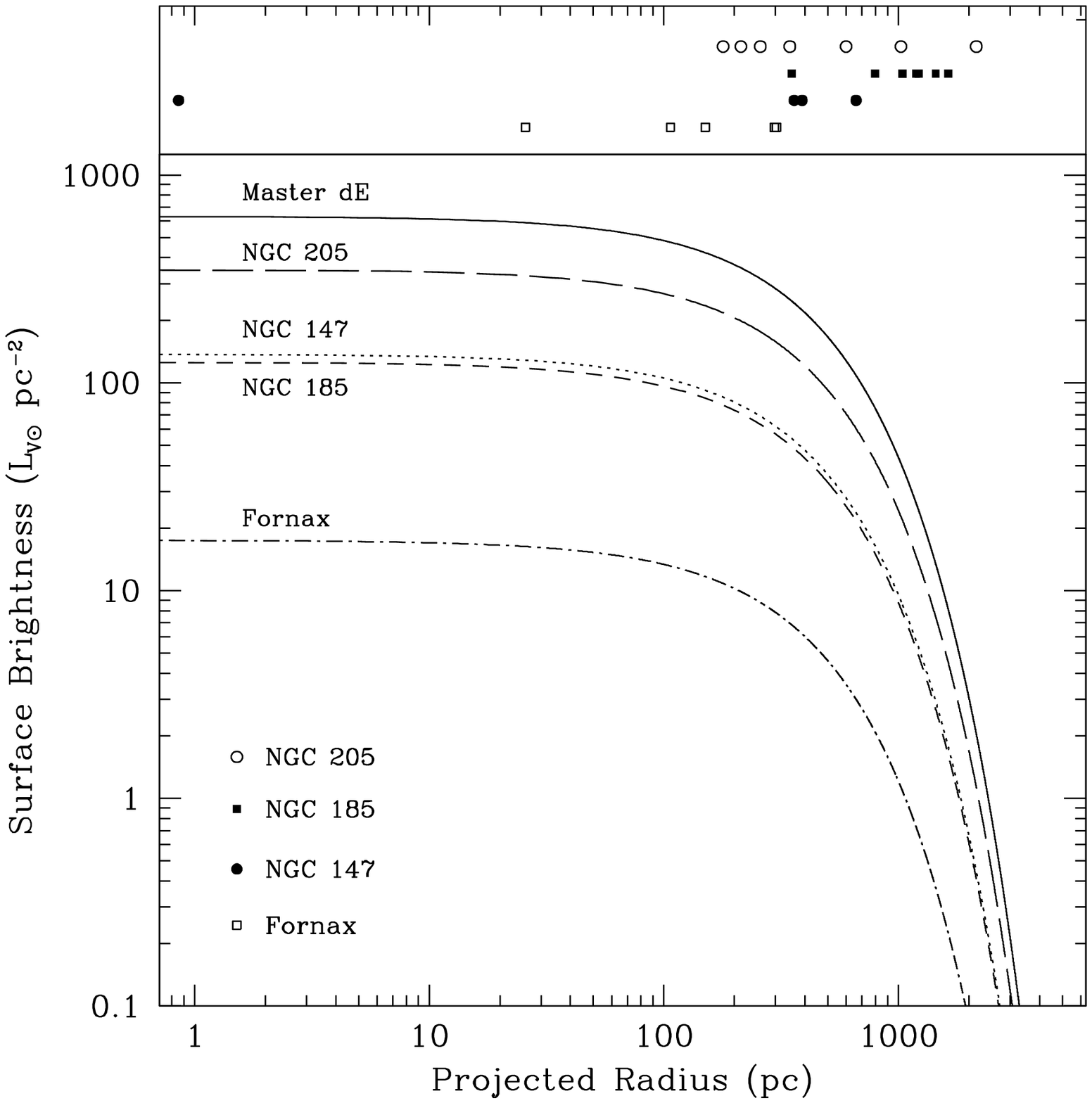}

\figcaption[cmw08.eps]{
Surface brightness profiles for the four dwarf galaxies 
used to construct the ``master dE". The curves represent exponential profiles
with scale lengths and central surface brightnesses taken from Caldwell et al. (1992). We
have corrected all four profiles to a common scale-length of $\alpha_s = 375$ pc before adding them
to create the master dE. The radial positions of the GCs in each galaxy, also corrected 
to a common scale-length, are indicated at the top of the figure.
\label{fig8}}

 
\plotone{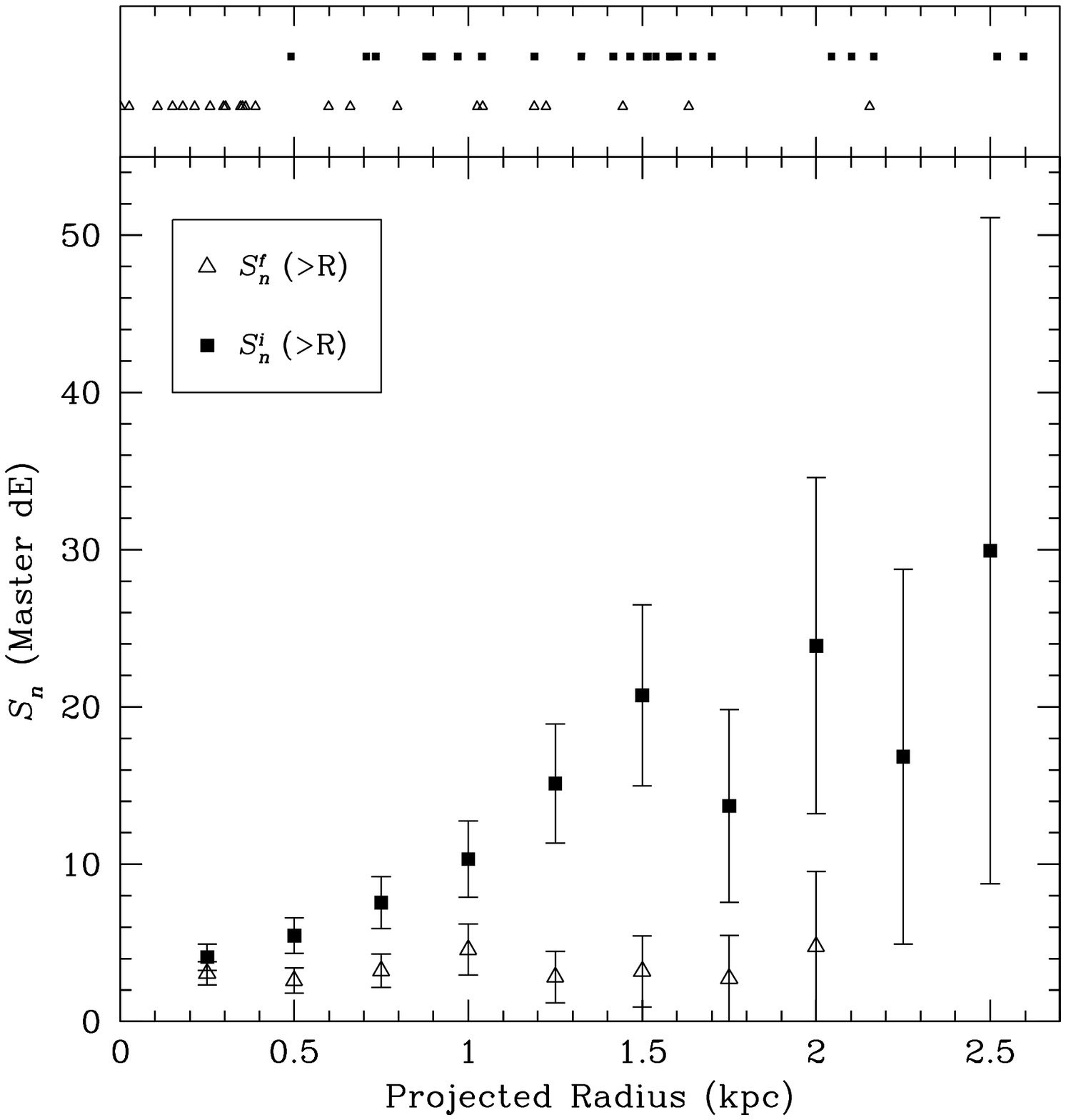}
 
\figcaption[cmw09.eps]{
Specific frequency profiles for the GCS of the master dE galaxy. The triangles 
indicate the present-day specific frequencies exterior to the marked positions. 
The squares indicates the same profile using the {\it initial} GCS distribution 
($i.e.$, after correcting the position of each GC
for the effects of dynamical friction). If placed in the core of the Virgo cluster,
a galaxy of this mass would be tidally stripped to a radius of $\sim$ 1 kpc. The
tidal debris would have $S_n$ $\gae$ 10, similar to that observed for the halo
of NGC 4486 (McLaughlin, Harris \& Hanes 1994). The initial and final GC positions 
are shown in the top panel by the filled and open symbols, respectively.
\label{fig9}}
 
 
\plotone{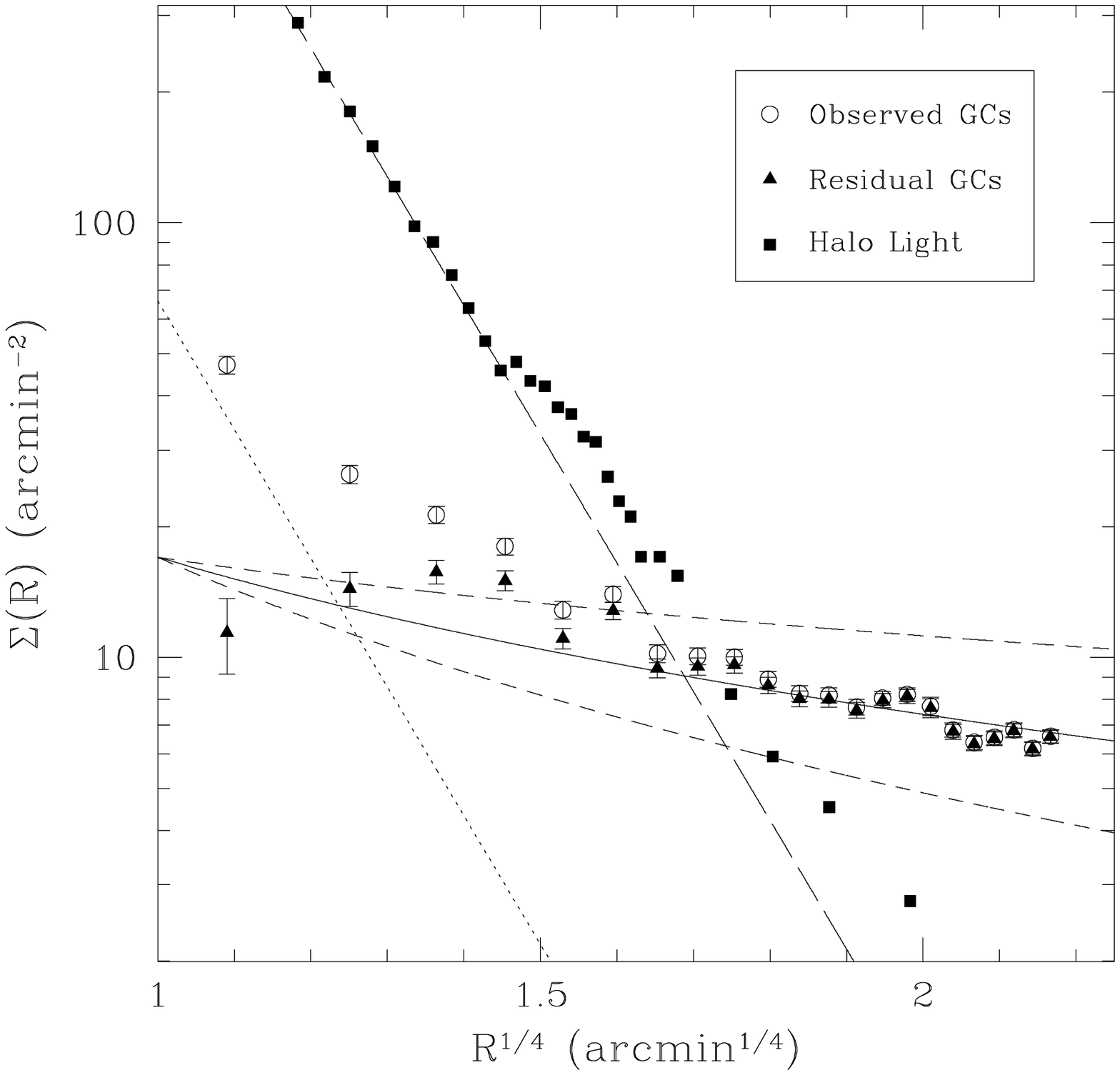}

\figcaption[cmw10.eps]{
Surface density profile for NGC 4486 GCs (open circles) taken from
the wide-field photographic survey of Harris (1986). 
Filled squares show the surface brightness profile of
the underlying halo light, arbitrarily shifted in the vertical direction.
The long dashed line shows the best-fit $R^{1/4}$ law for the galaxy light interior to
$R$ = 4\farcm5. Both the data and the best-fit profile are taken from de Vaucouleurs \& Nieto
(1978). The profile expected for the population of metal-rich GCs, should they
follow the same radial distribution as the stars themselves, is shown by
the dotted line. The zero-point has been determined from the relative numbers of 
metal-rich and metal-poor GCs measured by
Elson \& Santiago (1996) at a distance of 2\farcm5 from the galaxy's center
(see text for details). The filled triangles show the residuals between the
observed GC profile and that assumed for the metal-rich clusters.
The solid line shows the surface
density profile expected for a population of objects which trace the total mass
in the Virgo core, as measured by Cohen \& Ryzhov (1997). The dotted lines
represent the 1$\sigma$ confidence limits on the slope of the inferred density
profile: $\rho(r) \propto r^{\alpha}$ where $\alpha = 1.7\pm0.15$.
\label{fig10}}

 
\end{document}